\documentclass[preprint]{revtex4-1}
\usepackage{longtable}
\usepackage{latexsym}
\usepackage{graphics}
\usepackage{epsfig}
\usepackage{amsmath}
\usepackage{amssymb}
\usepackage{fancybox}
\usepackage{bm}
\usepackage{amsthm}
\usepackage{mathrsfs}
\usepackage{fancyhdr}
\usepackage{ascmac}

\def\mrmTD{\mathrm{TD}}

\def\iII{{I\hspace{-.1em}I\hspace{.1em}}}
\def\iIII{{I\hspace{-.1em}I\hspace{-.1em}I\hspace{.05em}}} 
\def\iIV{{I\hspace{-.2em}V\hspace{.05em}}}

\def\bullet{{a}}
\def\bullett{{ab}}

\def\www{{\mathfrak w}}

\def\ooo{\mathfrak{o}}
\def\OOO{\mathfrak{O}}

\def\r{{\bm r}}

\def\QQQ{\mathfrak{Q}}

\def\xxx{\mathfrak{x}}

\def\zzz{\mathfrak{z}}

\def\TTT{{\mathfrak T}}

\def\td{d_{\mathfrak w}}

\def\vvv{\mathfrak{v}}

\def\GGG{\mathfrak{G}}

\def\SSS{{\mathfrak S}}

\def\LLL{{\mathfrak L}}
\def\LD{\pounds}

\def\JJJ{{\mathfrak J}}

\def\HH{{\cal H}}

\def\HHH{{\mathfrak H}}

\def\aaa{{\mathfrak a}}

\def\BBB{{\mathfrak B}}
\def\bbb{{\mathfrak b}}

\def\CCC{{\mathfrak C}}
\def\ccc{{\mathfrak c}}

\def\Varepsilon{{\mathcal E}}
\def\EEE{\mathfrak{E}}
\def\eee{{\mathfrak e}}

\def\I{\mathscr{I}}

\def\M{{\cal M}}
\def\MM{\mathscr{M}}
\def\MMM{\mathfrak{M}}

\def\RRR{\mathfrak{R}}

\def\vomega{{\mathfrak w}}
\begin{document}
\title{
Geometrothermodynamics for Black holes and de Sitter Space
}
\author{Yoshimasa Kurihara}
\email{yoshimasa.kurihara@kek.jp}
\affiliation{The High Energy Accelerator Organization (KEK), 
Tsukuba, Ibaraki 305-0801, Japan}
\begin{abstract}
A general method to extract thermodynamic quantities from solutions of the Einstein equation is developed.
In 1994, Wald established that the entropy of a black hole could be identified as a Noether charge associated with a Killing vector of a global space-time (pseudo-Riemann) manifold.
We reconstruct Wald's method using geometrical language, e.g$.$, via differential forms defined on the local space-time (Minkowski) manifold.
Concurrently, the abstract thermodynamics are also reconstructed using geometrical terminology, which is parallel to general relativity.
The correspondence between the thermodynamics and general relativity can be seen clearly by comparing the two expressions.
This comparison requires a modification of Wald's method.

The new method is applied to Schwarzschild, Kerr, and Kerr--Newman black holes and de Sitter space.
The results are consistent with previous results obtained using various independent methods.
This strongly supports the validity of the area theorem for black holes.
\end{abstract}
\pacs{23.23.+x, 56.65.Dy}
\keywords{General relativity; Thermodynamics; Black holes}
\maketitle
\section{Introduction}\label{intro}
Some gravitational phenomena are known to have thermodynamic interpretations.
It was first pointed out by Bekenstein\cite{PhysRevD.7.2333} that black holes have entropy and follow the second law of thermodynamics.
After the pioneering work of Bekenstein and Hawking\cite{hawking1975}, intensive discussions revealed thermodynamical aspects of black holes.
Now, four laws of black hole mechanics\cite{Bardeen} are widely accepted:
The zeroth law states that the thermodynamic quantities, such as the temperature and surface gravity, remain constant over the horizon of a black hole\cite{PhysRevLett.26.331,WALD9507055}.
The first law can be expressed as\cite{Bardeen}
\begin{eqnarray}
dM_\mathrm{H}&=&\left(\kappa c/8\pi G\right)dA_\mathrm{H}+\Omega_\mathrm{H}
~dJ+\Phi~dQ,\label{th1st}
\end{eqnarray}
where $M$ is the mass of a black hole, $\kappa$ is the surface gravity, $A_\mathrm{H}$ is the area of the event horizon, $G$ is Newton's gravitational constant, $\Omega_\mathrm{H}$ is the surface angular velocity, $J$ is the angular momentum, $\Phi$ is the electromagnetic potential, and $Q$ is a charge.
For (\ref{th1st}) to be the first law of thermodynamics for a black hole, the relationship between black hole quantities, $\kappa,A_\mathrm{H}$, and the thermodynamic quantities, temperature and entropy, must be clarified. 
The temperature of a black hole can be equated with the Hawking temperature\cite{hawking1975}, $T_\mathrm{H}=\hbar\kappa/2\pi c$, which is obtained using a semi-classical treatment of the particle radiation from the surface of a black hole.
The entropy can be obtained from the area theorem\cite{hawking1975} for black holes:
$
S_\mathrm{H}={A_\mathrm{H}}/{4l_p^2}
$,
where $S_\mathrm{H}$ is the entropy of a black hole and $l_p=\sqrt{G\hbar/c^3}$ is the Planck length.
Using these relations, the first term of the r.h.s. of (\ref{th1st}) can be rewritten as
$
{\kappa c}/{(8\pi G)}dA_\mathrm{H}=T_\mathrm{H}dS_\mathrm{H}
$,
which allows a thermodynamic interpretation of (\ref{th1st}).
The area theorem is obtained in multiple ways for various types of black holes\cite{PhysRevD.35.449,PhysRevD.48.R3427,PhysRevD.50.846,PhysRevD.52.4430}.
Wald proposed equating the Noether charge with the entropy of black holes\cite{PhysRevD.48.R3427}.
Wald's method is applicable to a wide variety of Einsteinian and post-Einsteinian gravitational theories\cite{PhysRevD.61.084027}.
The black hole entropy has been discussed from both classical and quantum approaches\cite{THOOFT1985727,  PhysRevD.34.373, PhysRevD.47.1420, PhysRevD.49.6587, CALLAN199455, HOLZHEY1994443, Liberati:1996kt, PhysRevLett.80.904, Carlip:2007qh, 1126-6708-2009-11-109, PhysRevD.85.064031, COMPERE2015443, Compare2015, Saravani:2012is, PhysRevD.91.024032}, and also string theoretically\cite{STROMINGER199699, LUNIN2002342, PhysRevLett.96.181602, Hubeny:2007xt, Compare2016}.
A nice overall review article can also be found\cite{Wald2001}.
Gour and Mayo demonstrated that the entropy of a black hole is (nearly) linear in the area of the horizon\cite{PhysRevD.63.064005}.
The relation between black hole entropy and the Noether charge is discussed by many authors\cite{Fatibene:1998rq,Carlip:1999cy,Brustein:2009wr,Aros:2010jb,Majhi:2012tf,Chakraborty:2015hna,Setare:2015nla,Jacobson:2015uqa}.
In addition to those studies, the importance of boundary terms plays the essential role of the black hole entropy, that is first pointed out by Yoke\cite{PhysRevLett.28.1082}.
Especially, the necessity of null boundary is discussed by Parattu\cite{Parattu:2015gga,Parattu:2016trq}.
Interesting discussions related the boundary terms can be found in refs.\cite{Neiman:2012fx,Neiman:2013ap,Neiman:2013lxa,Jubb:2016qzt,Lehner:2016vdi,Krishnan:2016mcj}.
In spite of these intensive studies, the area theorem for a black hole entropy has not yet been proven generally. 

We propose a new method to extract entropy from the general relativistic theory.
Our method is sufficiently general to discuss the thermodynamics of a wide variety of solutions of the Einstein equation with an event horizon.
The Lagrangian and Hamiltonian formalisms are given in a completely covariant way in Sections \iII.
In Section \iIII, the abstract thermodynamic system is rewritten using the terminology for a geometry that is parallel to the treatment for general relativity given in Sections \iII.
The new method is applied to black holes and the de Sitter space in Section \iIV, and then the area theorem is examined for these examples.
Finally, a summary is presented in Section V.
Physical constants are set to unity, e.g., $4\pi G=\hbar=c=1$, in this work unless otherwise noted.

\section{Vierbein formalism of general relativity}\label{sec4}
First, classical general relativity is geometrically re-formulated in terms of a vierbein formalism. 
The formalism and terminology in this section primarily (but not completely) follow Fr{\` e}\cite{fre2012gravity}.
The vierbein formalism of general relativity can be found in other textbooks\cite{ashtekar1988new, carroll2004spacetime, Padmanabhan:2010zzb, bojowald2011quantum, rovelli2014covariant, poisson2014gravity} too.

In our formalism\cite{2017arXiv170305574K}, a spin form $\www^\bullett$ and a surface form $\SSS_\bullett$ are treated as fundamental phase-space variables.
The spin- and surface-forms are respectively defined in a local Lorentz manifold $\M$ as $\www^{a_1a_2}=\omega_{\mu~~b}^{~a_1}\eta^{b a_2}dx^\mu$ and $\SSS_{a_1a_2}=\epsilon_{a_1a_2b_1b_2}\Varepsilon^{b_1}_\mu\Varepsilon^{b_2}_\nu dx^\mu\wedge dx^\nu/2$, where $\omega_{\mu~a_2}^{~a_1}$ is a spin connection and $\Varepsilon^a_\mu$ is a vierbein.
We use a local Lorentz metric $\eta_\bullett=diag(1,-1,-1,-1)$ and a Levi Civita tensor (symbol)  $\epsilon_{0123}=1$ on the local Lorentz manifold $\M$.
A torsion two-form $\TTT^\bullet$ can be defined as $\TTT^\bullet=\td\eee^\bullet$ using a local $SO(1,3)$ covariant derivative $\td$.
In this work, the Fraktur letters are used for differential forms, and Greek- and Roman-indices, both run from zero to three,  are used for a coordinate on the global and local space time manifold, respectively.
A curvature two-form and a volume from are introduced as $\RRR^{a_1a_2}=d\www^{a_1a_2}+\www^{a_1}_{~~b}\wedge\www^{ba_2}$ and $\vvv=\epsilon_{a_1a_2a_3a_4}\eee^{a_1}\wedge\eee^{a_2}\wedge\eee^{a_3}\wedge\eee^{a_4}/4!$, respectively.
The Einstein--Hilbert gravitational Lagrangian is expressed using these forms as;
\begin{eqnarray}
\LLL_G
&=&\frac{1}{2}\left(
\RRR^{a_1a_2}\wedge\SSS_{a_1a_2}
-\frac{\Lambda}{3!}\vvv
\right),\label{Lagrangian44}
\end{eqnarray}
where $\Lambda$ is the cosmological constant.
The Einstein equation and torsionless condition can be obtained as a Euler-Lagrange equation  by taking a variation with respect to the surface and spin forms, respectively.

Next, we rewrite the classical gravitation theory in the Hamiltonian formalism in a completely covariant way.
The gravitational Lagrangian form (\ref{Lagrangian44}) can be represented as
\begin{eqnarray}
\LLL_G&=&
\frac{1}{2}\RRR^{a_1a_2}\wedge\SSS_{a_1a_2}
+\frac{1}{4!}\Lambda~\epsilon^{a_1a_2a_3a_4}\SSS_{a_1a_2}\wedge\SSS_{a_1a_2},
\nonumber\\
&=&\frac{1}{2}\left(
d\vomega^{a_1a_2}
+\vomega^{a_1}_{~~b}\wedge\vomega^{ba_2}
-\frac{\Lambda}{3!}\overline{\SSS}^{a_1a_2}\right)\wedge{\SSS}_{a_4a_3},\label{Lagrangian05-1}
\end{eqnarray}
where $\epsilon^{a_1a_2a_3a_4}=-\epsilon_{a_1a_2a_3a_4}$ is used.
Here, a new operator, called the {\it $\epsilon$-conjugate operator}, is introduced for shorter representations including Levi-Civita tensors.
This is a map that maps a rank-$p$ tensor field to a rank-$(n-p)$  tensor field defined on an $n$-dimensional space-time manifold such that
\begin{eqnarray*}
\overline{\aaa^{a_1\cdots a_p}}&=&\overline{\aaa}_{a_1\cdots a_{n-p}}=~~
\frac{1}{p!}\epsilon_{b_1\cdots b_p a_1\cdots a_{n-p}}
\aaa^{b_1\cdots b_p},\\
\overline{\aaa_{a_1\cdots a_p}}&=&\overline{\aaa}^{a_1\cdots a_{n-p}}=-\frac{1}{p!}
\epsilon^{a_1\cdots a_{n-p} b_1\cdots b_p}
\aaa_{b_1\cdots b_p},
\end{eqnarray*}
where $\aaa$ is an arbitrary form of a rank-$p$ tensor and $\epsilon$ is a completely anti-symmetric tensor in $n$-dimensions.
Note that $\overline{\overline{\aaa}}=\aaa$ when $\aaa$ is anti-symmetric with respect to all indices. 
The surface form can be expressed using the $\epsilon$-conjugate operator as
$\SSS_{ab}=\overline{\eee^a\wedge\eee^b}$.
Therefore, the $\epsilon$-conjugate of the surface form, which appears frequently in later parts of this report, is $\overline{\SSS}^{ab}=\overline{\SSS_{ab}}={\eee^a\wedge\eee^b}$. 
In the four-dimensional manifold, the identities
 \begin{eqnarray*}
\aaa^{a_1}\wedge\overline{\ccc}_{a_1}& =&\frac{1}{3!}\overline{\aaa}_{c_1c_2c_3}\wedge\ccc^{c_1c_2c_3},\\
\bbb^{b_1b_2}\wedge\overline{\BBB}_{b_1b_2}&=&
\overline{\bbb}_{b_1b_2}\wedge{\BBB}^{b_1b_2}
\end{eqnarray*}
 are satisfied for any rank-1 tensor $\aaa^{a_1}$, rank-2 tensors $\bbb^{b_1b_2}$ and $\BBB^{b_1b_2}$, and rank-3 tensor $\ccc^{c_1c_2c_3}$.
Note that the $\epsilon$-conjugate is not the same as the Hodge-star operator, whose definition includes the metric tensor.
One has to avoid to use the Hodge-star operator to construct a gravitational theory because the metric tensor can be obtained only after solving the Einstein equation.
 
For the Lagrangian density of the gravity, it is impossible to specify which terms are kinetic and potential energies.
This is reasonable because the gravitational energy cannot be defined locally in general relativity. 
Moreover, even on a local Minkowski manifold, the discrimination between kinetic and potential energies is frame dependent.
For example, no gravitational (potential) energy exists in the local inertial frame.
Start from the Lagrangian density form (\ref{Lagrangian05-1}):
\begin{eqnarray}
{\LLL}_G
&=&\frac{1}{2}\left(
\GGG_\Lambda^{a_1a_2}(\vomega,\SSS)+\frac{\Lambda}{3!}\overline{\SSS}^{a_1a_2}
\right)
\wedge{\SSS}_{a_1a_2},\label{Lagrangian05}
\end{eqnarray}
where
\begin{eqnarray*}
\GGG_\Lambda^{a_1a_2}(\vomega,\SSS)&=&
d\vomega^{a_1a_2}
+\vomega^{a_1}_{~~b}\wedge\vomega^{ba_2}
-\frac{2\Lambda}{3!}\overline{\SSS}^{a_1a_2}.
\end{eqnarray*}
Using this two-form, the Einstein equation can be expressed as 
\begin{eqnarray}
\epsilon_{\bullet b_1b_2b_3}\GGG_\Lambda^{b_1b_2}\wedge\eee^{b_3}=0.\label{EoM3}
\end{eqnarray}

Under these conditions, a covariant Hamiltonian formalism is introduced as follows.
First, $\vomega$ is identified as the first canonical variable.
Then, the second canonical variable $\MMM$ (the canonical momentum) is introduced as 
\begin{eqnarray}
\MMM_{a_1a_2}=\frac{\delta{\LLL}_G}{\delta\left(d\vomega^{a_1a_2}\right)}=\SSS_{a_1a_2}.
\end{eqnarray}
The reason why the spin form is taken as the first canonical variable will be discussed in Section \ref{sec6}.
Therefore, the two fundamental forms $\{\vomega,\MMM=\SSS\}$ form a generalized phase space.
The Hamiltonian density can be obtained as
\begin{eqnarray}
\HHH_G&=&
\frac{1}{2}d\vomega^{a_1a_2}\wedge\MMM_{a_1a_2}-
{\LLL}_G,\nonumber\\
&=&
\frac{1}{2}\left(
-\vomega^{a_1}_{~~b}\wedge\vomega^{ba_2}
+\frac{\Lambda}{3!}\overline{\SSS}^{a_1a_2}
\right)\wedge\SSS_{a_1a_2}.\label{HG}
\end{eqnarray}
Note that this Hamiltonian density does not contain any information concerning the dynamics (the derivatives of the fields).
The canonical equations can then be obtained immediately:
\begin{eqnarray}
\epsilon_{\bullet b_1b_2b_3}\frac{\delta\HHH_G}{\delta\MMM_{b_1b_2}}\wedge\eee^{b_3}&=&
\epsilon_{\bullet b_1b_2b_3}d\vomega^{b_1b_2}\wedge\eee^{b_3}.\label{canonicaleq1}\\
\frac{\delta\HHH_G}{\delta\vomega^{\bullet b}}\wedge\eee^a&=&-d\MMM_{\bullet b}\wedge\eee^a.\label{canonicaleq2}
\end{eqnarray}
The first equation of (\ref{canonicaleq1}) gives the equation of motion:
\begin{eqnarray}
\left(\frac{1}{2}\RRR^{a_1a_2}-
\frac{\Lambda}{3!}\overline{\SSS}^{a_1a_2}\right)\wedge\SSS_{a_1a_2}&=&0,
\end{eqnarray}
which simply leads to the Einstein equation, $(\ref{EoM3})$, and the second equation of (\ref{canonicaleq1}),
\begin{eqnarray}
\td\SSS_{\bullet b}\wedge\eee^a&=&2\TTT^a\wedge\SSS_{\bullet b}=0,
\end{eqnarray}
leads to the torsionless condition, as expected. 
\section{Geometrothermodynamic formalism}\label{sec5}
In this section, before discussing the thermodynamic aspects of the gravitation, the thermodynamic system is rewritten using geometrical terminologies.
The formalization is based on that of ref.\cite{kurihara2014}; however, it is rewritten according to the conventions of this work.
Here, we confine our attention to the formal relationships between the thermodynamic variables and ignore their physical meanings.
 
A characteristic function of the thermodynamics can be introduced as
\begin{eqnarray}
f_{\mrmTD}&=&-TS+pV+\mu N.
\end{eqnarray}
Connotation of each variables is that $S$ is the entropy, $T$ is the temperature, $p$ is the pressure, $V$ is the volume, $N$ is the number of molecules, and $\mu$ is the chemical potential.
These variables can be categorized into two types: extensive and intensive variables.
According to the natural connotation given above, $\xi^a=(S,V,N)$ are categorized as extensive variables and $\zeta^a=(T,p,\mu)$ are intensive variables, where $a=0,1,2$.
We assume the existence of a smooth three-dimensional base-manifold $\M_\mrmTD$ and $\xi$  and $\zeta$ are vectors belonging to the tangent and cotangent bundles on the manifold, respectively.

Even though the thermodynamics can be constructed starting from either $\xi^a$ or $\zeta^a$, here an extensive set is considered to be the first canonical variables (the coordinate space).
Therefore, the intensive variables are considered to be functions of the extensive variables.
According to ref.\cite{kurihara2014}, $S$ is considered to be the order parameter of the system and  $T=\partial f_{\mrmTD}/\partial S$ is considered as the ``Hamiltonian'' (with a narrower meaning compared with the discussions in previous sections that induce a ``time'' evolution in the system).
The intensive variables are not independent of each other; however, $T$ is assumed to be  independent of $S$.
Therefore, we can write the intensive variables as
\begin{eqnarray}
\left\{
\begin{array}{l}
T=T(V,N,p,\mu),\\
p=p(S,V,N),\\
\mu=\mu(S,V,N). 
\end{array}
\right.
\end{eqnarray}
The characteristic one-form is introduced as
\begin{eqnarray}
\vomega_\mrmTD&=&\eta_{a_1a_2}\zeta^{a_1}\xxx^{a_2}=\zeta_a\xxx^a,\nonumber\\
&=&-TdS+pdV+\mu dN,
\end{eqnarray}
where the metric tensor is chosen to be $\eta_{ab}=\mathrm{diag}(-1,1,1)$ and the one-form is defined as  $\xxx^a=d\xi^a$.
The characteristic one-form can be thought of as an internal energy $\vomega_\mrmTD=-dU$.
This metric tensor suggests that the system has a $SO(1,2)$ symmetry and is consistent with an interpretation that the variable $S$ plays the role of ``time'' (i.e., an order  parameter) and ``T'' plays the role of an energy (i.e., the Hamiltonian). 
The characteristic two-form (the Lagrangian-density form) can also be written as
\begin{eqnarray}
\LLL_\mrmTD&=&d\vomega_\mrmTD
=\eta_{a_1a_2}\zzz^{a_1}\wedge\xxx^{a_2}=\zzz_a\wedge\xxx^a,\nonumber\\
&=&-dT\wedge dS+dp\wedge dV+d\mu \wedge dN,
\end{eqnarray}
where $\zzz^a=d\zeta^a$.
The Lagrangian form is a closed form because $d\LLL_\mrmTD=0$ from the definition. 
Therefore, a base manifold $\M_\mrmTD$ is a symplectic manifold with a symplectic potential of $\OOO_\mrmTD = \vomega_\mrmTD$.
The intensive variables are recognized as the canonical momentum:
\begin{eqnarray}
\zzz_\bullet&=&\frac{\delta\LLL_\mrmTD}{\delta\xxx^\bullet}=d\zeta_\bullet.
\end{eqnarray}
The action integral can be expressed as
\begin{eqnarray}
\I_\mrmTD&=&\int_{\Sigma^{(2)}}\LLL_\mrmTD
=\int_{\partial\Sigma^{(1)}}\vomega_\mrmTD,
\end{eqnarray}
where the integration region ${\Sigma^{(2)}}$ is an appropriate two-dimensional manifold embedded in the base manifold $\M$ and $\partial\Sigma^{(1)}$ is its boundary.
The equations of motion that represent the Maxwell relations of thermodynamics can be obtained in an explicitly coordinate-dependent manner.
The variational operation of the characteristic one-form with respect to $\xi^0=S$ is
\begin{widetext}
\begin{eqnarray}
\delta_S\vomega_\mrmTD&=&-\left(
\frac{\partial T}{\partial S}dS+
\frac{\partial T}{\partial p}dp+
\frac{\partial T}{\partial\mu}d\mu
\right)\delta S\nonumber+
\frac{dp}{dS}dV\delta S+
\frac{d\mu}{dS}dN\delta S-T\delta\left(dS\right),\\
&=&\left[
\left(
-\frac{\partial T}{\partial S}+\frac{dT}{dS}
\right)dS
+\left(
-\frac{\partial T}{\partial p}+\frac{dV}{dS}
\right)dp
+\left(
-\frac{\partial T}{\partial\mu}+\frac{dN}{dS}
\right)d\mu
\right]\delta S-d(T\delta S),\nonumber\\&=&0.
\end{eqnarray}
\end{widetext}

Therefore, the equations of motion are
\begin{eqnarray}
\frac{dT}{dS}=\frac{\partial T}{\partial S}=0,~~~\frac{dV}{dS}=\frac{\partial T}{\partial p},~~~
\frac{dN}{dS}=\frac{\partial T}{\partial\mu},
\end{eqnarray}
where the surface integration vanishes without applying any boundary condition because $\partial(\partial\Sigma^{(1)})=0$ (a boundary of a boundary does not exist).
Other sets of Maxwell relations can be obtained using variations with respect to $V$ or $N$.
The above equations represent the canonical equation of motion with the following interpretations: $S \leftrightarrow t$ (time) and $T\leftrightarrow\HH$ (Hamiltonian).
These equations can also be written as the Euler--Lagrange equation of motion in a coordinate-independent manner as $\delta_\xxx\I_\mrmTD=0\Rightarrow \zzz^\bullet=0$, which is, of course, equivalent to the canonical equations.
The surface term
$
\OOO_\delta=\zeta_a~\delta\xxx^a
$
can be considered as another type of the symplectic potential.
\begin{table}[b]
\begin{center} 
\begin{tabular}{cccl}
\hline
Name& Type &~&~~Definition\\
\hline\hline
Symplectic potential& 1-form&$\OOO_\mrmTD$&$=~\vomega_\mrmTD
~=~\zeta_a\xxx^a$\\
%
%
%
Noether current & 1-form&
$\JJJ_\upsilon$&$=~\OOO_\upsilon-\iota_\upsilon\LLL_\mrmTD$\\
Noether charge &0-form &
$\QQQ_\upsilon$&$=~\zeta_a\iota_\upsilon\xxx^a$\\
\hline
\end{tabular}
\caption{
Symplectic potential and the Noether current/charge in thermodynamics.
 }\label{formsE}.
\end{center}
\end{table}

Next, consider an infinitesimal translation of the characteristic one-form along a vector field $\upsilon^a$ on the base manifold $\M_\mrmTD$ such that
\begin{eqnarray}
\delta_\upsilon\LLL_\mrmTD&=&\LD_\upsilon\LLL_\mrmTD
=(\LD_\upsilon\zzz_a)\wedge\xxx^a+
(d\zeta_a)\wedge\LD_\upsilon\xxx^a,\nonumber\\
&=&(\LD_\upsilon\zzz_a)\wedge\xxx^a+d\OOO_\upsilon,\label{E1}
\end{eqnarray}
where $\LD_\upsilon$ is the Lie derivative along the vector $\upsilon^a$ and $\OOO_\upsilon = \zeta_a \LLL_\upsilon \xxx^a$ can be considered as yet another type of the symplectic potential.
Here $d\xxx^\bullet = d\zzz^\bullet = \iota_\upsilon\xi^\bullet = \iota_\upsilon\zeta^\bullet = 0$ are used.
The Noether current according to this translation symmetry can be obtained using $\OOO_\upsilon$ such that
\begin{eqnarray}
\JJJ_\upsilon&=&\OOO_\upsilon-\iota_\upsilon\LLL_\mrmTD.
\end{eqnarray}
It is confirmed that the Noether current is conserved via the explicit calculation of
\begin{eqnarray}
d\JJJ_\upsilon&=&
d\OOO_\upsilon-\LD_\upsilon\LLL_\mrmTD+\iota_\upsilon(d\LLL_\mrmTD),\\
&=&-(\LD_\upsilon\zzz_a)\wedge\xxx^a~=~0,
\end{eqnarray}
due to (\ref{E1}), $d\LLL_\mrmTD=0$, and the equation of motion $\zzz^\bullet=0$.
Therefore, the Noether current can be represented using a closed form as $\JJJ_\upsilon = d\QQQ_\mrmTD+\CCC$, where $\QQQ_\mrmTD$ is the Noether charge and $\CCC$ is a term that is zero for the solution of the equation of motion.
We can see that
\begin{eqnarray}
\JJJ_\upsilon
&=&d(\zeta_a\iota_\upsilon\xxx^a)
-\zzz_a\iota_\upsilon\xxx^a-
\iota_\upsilon(\zzz_a\wedge\xxx^a),\nonumber \\
&=&d(\zeta_a\iota_\upsilon\xxx^a)
-\iota_\upsilon\zzz_a\xxx^a,
\end{eqnarray}
where the second term is zero for the solution of the equation of motion.
Therefore, the Noether charge can be extracted as follows:
\begin{eqnarray}
\QQQ_\upsilon&=&\zeta_a\iota_\upsilon\xxx^a.
\end{eqnarray}
The symplectic potential and the Noether current/charge are summarized in Table \ref{formsE}.

For the case of the adiabatic free-expansion of an isolated gas, in which the temperature remains constant, the meaning given above is the actual physical case.
The vector $\upsilon^a=(1,0,0)$ is the ``Killing vector'' in this context because $\upsilon_a~\zzz^a=-dT=0$.
Therefore, the Noether charge can be singled out as 
\begin{eqnarray}
Q_\mrmTD&=&\eta_{00}\zeta^0\xi^0~=~-TS.\label{thNC}
\end{eqnarray}
In this case, the Noether charge is not simply the entropy; instead, it is the temperature times the entropy. 
\section{The symplectic formalism of general relativity}\label{sec6}
A general formalization of the thermodynamics of space-time is formalized in this section.
The method proposed in Wald\cite{PhysRevD.48.R3427} and Iyer and Wald\cite{PhysRevD.50.846,PhysRevD.52.4430} are carefully rewritten using our terminology.  

\subsection{General formalism}
Let us start from the Lagrangian density of (\ref{Lagrangian05}).
This Lagrangian form has two candidates for the first canonical variables. 
According to the geometrothermodynamic formulation introduced in the previous section, we would like to start from the intensive variables. The two-dimensional surface form $\SSS_{ab}$ can be easily recognized as an extensive variable. Conversely, the spin form $\vomega^{ab}$ may be intensive because it is defined locally, independent of the total volume under consideration.
Therefore, we identify the spin form as the first canonical variable, as discussed in Section \ref{sec4}.

Taking a variation of the Lagrangian density with respect to the first canonical variable $\vomega^{ab}$, one gets
\begin{eqnarray}
\delta_\vomega{\LLL}_G&=&
\delta\vomega^{a_1a_2}\wedge\overline{\EEE}_{a_1a_2}
+d\OOO_\delta,\label{Lvar}
\end{eqnarray}
where $\EEE^{a_1a_2}=2\TTT^{a_1}\wedge\eee^{a_2}$ and $\OOO_\delta=\delta\vomega^{a_1a_2}\wedge\SSS_{a_1a_2}$.
The second term, $d\OOO_\delta$, is the surface term.
When a surface term is set to zero due to the boundary condition, Hamilton's principle leads to the equation of motion $\EEE^{ab}=0$, which corresponds to the torsionless condition $\TTT^a=0$. 
Here, we introduce a three-form 
\begin{eqnarray}
\OOO&=&\frac{1}{2}\vomega^{a_1a_2}\wedge\SSS_{a_1a_2},
\end{eqnarray}
which is called the ``{\it symplectic potential}''.
Here, we assume the existence of an inverse of the spin form. 
Compared with the geometrothermodynamics given in the previous section, it is identified as the symplectic potential.
Using  this symplectic potential, the surface term can be expressed as $\OOO_\delta=\delta_\vomega\OOO=\delta\vomega^{a_1a_2}\wedge\SSS_{a_1a_2}$. 

Next, consider the Lie derivative of the Lagrangian form with respect to a vector field $X$ defined on the global manifold.
The Lie derivative along the vectors $\xi^\mu\in X$ of the Lagrangian density becomes
\begin{eqnarray}
\LD_\xi{\LLL}_G&=&\frac{1}{2}\LD_\xi
\left(
\GGG_\Lambda^{a_1a_2}+\frac{\Lambda}{3!}\overline{\SSS}^{a_1a_2}
\right)
\wedge\SSS_{a_1a_2}\nonumber\\
&~&+\frac{1}{2}
\left(
\GGG_\Lambda^{a_1a_2}+\frac{\Lambda}{3!}\overline{\SSS}^{a_1a_2}
\right)
\wedge\LD_\xi\SSS_{a_1a_2}.\label{LLiedel}
\end{eqnarray}
The Lie derivative and the external derivative are commutable with each other,
$
\LD_\xi{d}\vomega^{a_1a_2}={d}\LD_\xi\vomega^{a_1a_2},
$
by definition.
After simple manipulations, we finally arrive at the expression from (\ref{LLiedel}) without assuming the equations of motion:
\begin{eqnarray}
\LD_\xi{\LLL}_G&=&
d\OOO_\xi+
\frac{1}{2}\left(
\LD_\xi\vomega^{a_1a_2}\wedge\overline{\EEE}_{a_1a_2}
+\GGG_\Lambda^{~a_1a_2}\wedge\LD_\xi\SSS_{a_1a_2}\right),
\end{eqnarray}
where 
\begin{eqnarray}
\OOO_\xi&=&\frac{1}{2}\LD_\xi\vomega^{a_1a_2}\wedge\SSS_{a_1a_2}.
\end{eqnarray}
If solutions to the equations of motion are assumed, the Lie derivative of the Lagrangian density vanishes up to the total derivative of the function $\OOO_\xi$.
This symmetry may ensure the existence of a conserved current such as
\begin{eqnarray}
\JJJ_\xi&=&
\OOO_\xi-\iota_\xi\LLL_G,\label{Noether}
\end{eqnarray}
which is known as the Noether current.
Here, $\iota_\xi$ is a contraction between the form and the vector.
Therefore, the second term of  (\ref{Noether}) is a three-form as well as the first term.
The conservation of the Noether current can be confirmed via
\begin{eqnarray}
d\JJJ_\xi&=&d\OOO_\xi-\LD_\xi\LLL_G+
\iota_\xi\left(d\LLL_G\right),\nonumber\\
&=&-\frac{1}{2}\left(
\LD_\xi\vomega^{a_1a_2}\wedge\overline{\EEE}_{a_1a_2}
+\GGG_\Lambda^{~a_1a_2}\wedge\LD_\xi\SSS_{a_1a_2}\right)
=0,\label{Noethercons}
\end{eqnarray}
where $d\LLL_G=0$ and two equations of motion are used.

Next, let us introduce a generating function, $\HH_\xi$, and its density form, $\HHH_\xi$, such that $\HH_\xi=\int\HHH_\xi$, where the integration region is the appropriate three-dimensional manifold embedded in the global space-time manifold. 
$\HHH_\xi$ can be understood as a generalization of the Hamiltonian introduced in the previous section, which induces a flow in the system along the general vector fields on the local manifold.
For example, when a time-like vector field can be defined globally in the local manifold and is chosen as the ``time coordinate'', the generating function density form $\HHH_t$ is just the standard Hamiltonian with its narrow meaning, which results in the ``time evolution'' of the system.

When the vector field $X$ is defined globally on the entire global manifold, the generating function with respect to the vector $\xi^\mu\in X$ can be defined in the follow manner.
Take the variation of the Noether current with respect to the spin form to be
\begin{eqnarray*}
\delta_\vomega\JJJ_\xi&=&\delta_\vomega\OOO_\xi-
\delta_\vomega(\iota_\xi\LLL_G),\nonumber\\
&=&
\LD_\xi\delta\vomega^{a_1a_2}\wedge\SSS_{a_1a_2}
-\LD_\xi\left(
\delta\vomega^{a_1a_2}\wedge\SSS_{a_1a_2}
\right)+d\iota_\xi\OOO_\delta,\nonumber\\
&=&
\ooo+d\iota_\xi\OOO_\delta,
\end{eqnarray*}
where $\ooo=  \LD_\xi\delta\vomega^{a_1a_2}\wedge\SSS_{a_1a_2} -\LD_\xi(\delta\vomega^{a_1a_2}\wedge\SSS_{a_1a_2})$.
Here, the equations of motion are used again.
The three-form  $\ooo$ is called the ``{\it symplectic current}''.
The generating function density form is introduced using the symplectic current so that $\delta_\vomega\HHH_\xi=\ooo$.
Therefore, the Noether current can be written as
\begin{eqnarray}
\delta_\vomega\JJJ_\xi&=&
\delta_\vomega\HHH_\xi+d\iota_\xi\OOO_\delta.\label{dj1}
\end{eqnarray}
Conversely, because the Noether current is the closed form for the solution of the equations of motion as shown in (\ref{Noethercons}), it can be expressed as
\begin{eqnarray}
\JJJ_\xi&=&d\QQQ_\xi+\CCC.\label{Noether2}
\end{eqnarray}
due to the Poincar\'{e} lemma, at least locally.
Here, the three-form $\CCC$ is eliminated using solutions of the equations of motion.
The two-form $\QQQ_\xi$ is the Noether charge of the gravitational field.
One can actually calculate the Noether current (\ref{Noether}) expressed in the shape of  (\ref{Noether2})  such that
\begin{eqnarray}
\JJJ_\xi&=&\frac{1}{2}
\LD_\xi\vomega^{a_1a_2}\wedge\SSS_{a_1a_2}-\iota_\xi\LLL_G,\nonumber\\
&=&\frac{1}{2}d\left(\left(\iota_\xi\vomega^{a_1a_2}\right)\SSS_{a_1a_2}\right)
-\frac{1}{2}\left(
(\iota_\xi\vomega^{a_1a_2})\bar{\EEE}_{a_1a_2}+
\GGG_\Lambda^{a_1a_2}\wedge\iota_\xi\SSS_{a_1a_2}
\right).\nonumber\\
\end{eqnarray}
Therefore, one can write the Noether charge as
\begin{eqnarray}
\QQQ_\xi&=&\frac{1}{2}\left(\iota_\xi\vomega^{a_1a_2}\right)\SSS_{a_1a_2}.\label{GRNC}
\end{eqnarray}
The integration of the equation (\ref{dj1}) can be expressed as
\begin{eqnarray}
\delta_\vomega\int_{\Sigma^{(3)}}\JJJ_\xi&=&
\delta_\vomega\HH_\xi+
\delta_\vomega\int_{\partial\Sigma^{(2)}}\iota_\xi\OOO,\label{Idj1}
\end{eqnarray}
where $\partial\Sigma^{(2)}$ is the boundary of $\Sigma^{(3)}$.
From  (\ref{Noether2}), (\ref{Idj1}), and the equations of motion, the generating function can be obtained as follows:
\begin{eqnarray}
\HH_\xi&=&\int_{\partial\Sigma^{(2)}}\left(
\QQQ_\xi-\iota_\xi\OOO
\right),\nonumber\\
&=&\int_{\partial\Sigma^{(2)}}
\vomega^{a_1a_2}\wedge\iota_\xi\SSS_{a_1a_2}.
\end{eqnarray}
The symplectic potential and Noether current/charge in general relativity are summarized in Table \ref{forms}.
\begin{table}[b]
\begin{center} 
\begin{tabular}{cccl}
\hline
Name& Type &~&~~Definition\\
\hline\hline
Symplectic potential& 3-form&$\OOO$&$=~\vomega^{a_1a_2}\wedge\SSS_{a_1a_2}/2$\\
%
%
%
%
Noether current & 3-form&
$\JJJ_\xi$&$=~\OOO_\xi-\iota_\xi\LLL_G$\\
Noether charge &2-form &
$\QQQ_\xi$&$=~(\iota_\xi\vomega^{a_1a_2})\SSS_{a_1a_2}/2$\\
%
\hline
\end{tabular}
\caption{
Symplectic potential and the Noether current/charge in general relativity.
 }\label{forms}
\end{center}
\end{table}

Based on the above geometrical formalization, one can now single the ``{\it entropy}'' out from the general relativity system by comparing the results in Sections \ref{sec5} and \ref{sec6}. 
In general relativity ($GR$), the first canonical variable is taken to be $\vomega^{\bullett}$, which may correspond to $\xi^\bullet=(S,V,N)$ in thermodynamics ($TD$).
The symplectic structure can be discussed in parallel for $TD$ and $GR$.
For example, using the surface term 
$\OOO_\mrmTD
\leftrightarrow\OOO_\delta
$,
the conserved Noether current can be introduced as $\JJJ_\upsilon\leftrightarrow\JJJ_\xi$.
The $TD$ analysis suggests that the entropy can be extracted from the Noether charge for $Q_\mrmTD\leftrightarrow\QQQ_\xi$.

Before closing this section, we would like to emphasize the origin to ensure the conservation of the Noether current/charge.
In above discussions, any specific properties of the vector fields are not used.
Though the Killing vector-fields will be takes as the vector filed $X$ in a next section, the conserved current/charge can be obtained for any vector filed.
The continuous symmetry to induce the Noethe\rq{}s theorem in this case is the variational operation with respect to the first canonical form $\vomega^{\bullett}$, which caused at most a term of total derivative as shown in (\ref{Lvar}).
Some examples of the Noether charge are summarized in {\bf Appendix A}.
Moreover concrete representations for the Lie derivative $\LD_\xi$ and contraction $\iota_\xi$ are not important.
Algebraic rules for differential forms play a essential role.

\subsection{Application to a Schwarzschild black hole}
Let us calculate the Noether charge for the Schwarzschild black hole solution, as our first example of this method.
The Schwarzschild solution is
\begin{eqnarray}
ds_\mathrm{Schw}^2&=&f^2_\mathrm{Schw}(r)dt^2-f^{-2}_\mathrm{Schw}(r)dr^2\nonumber\\
&~&~~~~-r^2\left(d\theta^2+\sin^2{\phi}~d\phi^2\right),\label{lmschw}
\end{eqnarray}
with a coordinate of $x^a=(t,r,\theta,\phi)$, where $f^2_\mathrm{Schw}(r)=1-2M/r$ and $M$ is the mass of a black hole.
The vierbein form can be extracted as
\begin{eqnarray}
\eee^\bullet&=&\left(
f_\mathrm{Schw}dt,
f_\mathrm{Schw}^{-1}dr,
r~d\theta,r\sin{\theta}~d\phi\right).\label{virschw}
\end{eqnarray}
From the above solutions and the torsionless condition, a unique solution of the spin form can be obtained\cite{fre2012gravity}:
\begin{eqnarray}
\vomega_\mathrm{Schw}^{\bullett}=\left(
\begin{array}{cccc}
0&-M/r^2~dt~& 0 & 0 \\
~& 0 &  f_\mathrm{Schw}d\theta &
f_\mathrm{Schw}\sin{\theta}~d\phi \\
~&~&0&\cos{\theta}~d\phi\\
~&~&~&0\\
\end{array}
\right),\nonumber\\
\end{eqnarray}
where the lower half is omitted because it is obvious due to the antisymmetry of the spin form.
The Schwarzschild space-time has
 the following Killing vectors:
\begin{eqnarray}
\left\{
\begin{array}{l}
\xi_1^\bullet=(~1,~0,~0,~0~),\\
\xi_2^\bullet=(~0,~0,~0,~1~),\\
\xi_3^\bullet=(~0,~0,\sin{\phi},~~\cot{\theta}\cos{\phi}~),\\
\xi_4^\bullet=(~0,~0,\cos{\phi},-\cot{\theta}\sin{\phi}~),
\end{array}
\right.
\end{eqnarray}
with the natural coordinate bases of $\{\partial_t, \partial_r, \partial_\theta, \partial_\phi\}$ on the tangent bundle of the global manifold $\MM$.
The first two Killing vectors are trivial because the vierbein forms do not include any functions of  $t$ and $\phi$.
The third and fourth vectors satisfy the Killing equation such that
\begin{eqnarray*}
\frac{1}{r^2\sin^2{\theta}}\frac{\partial \xi_{3,4}^2}{\partial\phi}+
\frac{1}{r^2}\frac{\partial \xi_{3,4}^3}{\partial\theta}=0.
\end{eqnarray*}
The second Killing vector is not independent of the third and fourth vectors because $\xi_2=\xi_{3,4}$ with ($\theta=\pm\pi/4,~\phi=0$).

Let us take a closer look at the first Killing vector $\xi_1^\bullet$, which is simply denoted as $\xi^\bullet$ hereafter. 
In the region $r>2M$, outside of the event horizon of a Schwarzschild black hole, the zeroth component of the vierbein form $\eee^0$ is globally a time-like vector.
Therefore, $t$ can be interpreted as the time coordinate in this region.
It is expected that the thermodynamic quantities are related to this Killing vector, because, for example, the entropy can only be well defined for the thermal-equilibrium state, which is static with respect to $t$. 
A dual vector of the Killing vector is given by
\begin{eqnarray*}
\xi_\mu&=&\eta_{a_1a_2}\varepsilon_\mu^{a_1}\varepsilon_\nu^{a_2}\xi^\nu
~=~\left(1-\frac{2M}{r},0,0,0\right),
\end{eqnarray*}
and therefore the normalization of the Killing vector is given as
\begin{eqnarray*}
\xi_\mu\xi^\mu&=&1-\frac{2M}{r}\rightarrow 1~~(r\rightarrow\infty),
\end{eqnarray*}
and is null on the event horizon of a black hole.
A contraction of the spin form with respect to the Killing vector can be calculated  as
\begin{eqnarray*}
\iota_\xi\vomega^{ab}&=&\omega_{~~\mu}^{ab}\iota_\xi dx^\mu
~=~\omega_{~~\mu}^{ab}\xi^\mu~=~\omega_{~~0}^{ab}\partial_t.
\end{eqnarray*}
Therefore, the Noether charge can be obtained as
\begin{eqnarray}
\frac{1}{k_\mathrm{E}}\int\QQQ_\xi&=&-\frac{1}{2k_\mathrm{E}}\int\frac{GM}{c^2}\sin{\theta}~d\theta\wedge d\phi=-\frac{Mc^2}{2}.
\end{eqnarray}
Here, we put the factor $k_\mathrm{E}=(4\pi G/c^4)$ back in front of the Lagrangian in the action integral and write the physical parameters $G,c$ and $\hbar$ explicitly.
By comparing this result with the thermodynamic result given in the previous section (\ref{thNC}), this quantity can be described using the thermodynamic variables as
 \begin{eqnarray}
\frac{Mc^2}{2}&=&S_\mathrm{Schw}T_\mathrm{Schw},
\end{eqnarray}
where $S_\mathrm{Schw}$ and $T_\mathrm{Schw}$ are the entropy and temperature of a Schwarzschild black hole, respectively.
A purely classical treatment of the gravity cannot resolve $S_\mathrm{Schw}$ and $T_\mathrm{Schw}$ unambiguously because the temperature of a black hole may appear as a quantum effect.
The semi-classical analysis of a black hole temperature by Hawking\cite{hawking1975} shows the temperature as
\begin{eqnarray}
T_\mathrm{H}&=&\frac{\hbar}{c}\frac{\kappa}{2\pi}.\label{howkingT}
\end{eqnarray}
We use this temperature to extract an expression of the entropy.
We note that this relationship between the temperature and the surface gravity is common to all event horizons.
Subsequently, the black hole entropy $S_\mathrm{Schw}$ can be obtained as
\begin{eqnarray}
S_\mathrm{Schw}&=&\frac{Mc^2}{2T_\mathrm{Schw}}
~=~\frac{M}{2}\left(
\frac{\hbar c}{8\pi GM}
\right)^{-1},\nonumber\\
&=&\frac{A_\mathrm{Schw}}{4l_p^2},
\end{eqnarray}
where $A_\mathrm{Schw}=4\pi(2MG/c^2)^2$ is the area of the event horizon of a black hole.
Here, the diameter of the event horizon of the hole $r_\mathrm{Schw}=2GM/c^2$ is used.
 A Hawking temperature of $T_\mathrm{Schw}=\hbar c^3/(8\pi GM)$ can be obtained from (\ref{howkingT}) and the surface gravity $\kappa=c^4/(4GM)$.
This entropy is equivalent to the celebrated Bekenstein--Hawking entropy \cite{PhysRevD.7.2333,hawking1975} of a Schwarzschild  black hole.

The differences between Wald's original method\cite{PhysRevD.48.R3427} and the method given here is as follows:
in Wald's method, the Killing vector was normalized such that the Killing field had a unit surface gravity.
This normalization results in a factor of $4GM/c^4$ in the Noether charge.
(Wald explained in \cite{PhysRevD.48.R3427} that this normalization makes the Noether charge local.)
In addition, Wald's definition of the entropy has a factor of $2\pi(c/\hbar)
$\footnote{The factor $c/\hbar$ was restored by the author to make the entropy dimensionless. This factor was not explicitly written (set to unity)  in Eq. (26) in the original paper\cite{PhysRevD.48.R3427}.} 
in front of the surface integration of the Noether charge.
Therefore, Wald's entropy has an additional factor of $8\pi GM/\hbar c^3$ compared with our definition, which corresponds to the inverse temperature of a black hole.
This is why the integration of the Noether charge simply gives the entropy in \cite{PhysRevD.48.R3427}. 
From a purely mathematical point of view, the two methods are equivalent to each other.

\subsection{Application to a Kerr--Newman black hole}
For the second example, we will treat the general solution of a Kerr--Newman black hole.
The line element of a black hole having a mass $M$, an angular momentum $J=\alpha M$, and an electric charge $q$ can be expressed using the Boyer--Lindquist metric\cite{BoyerLindquist} as\cite{PhysRevLett.11.237,Newman}

\begin{eqnarray}
ds_\mathrm{KN}^2&=&-\frac{\Delta}{\rho}\left(
dt-\alpha\sin^2{\theta}d\phi
\right)^2+\frac{\rho^2}{\Delta}dr^2+\rho^2d\theta^2
+\frac{1}{\rho^2}\sin^2{\theta}\left[
\left(r^2+\alpha^2\right)d\phi-\alpha dt
\right]^2,
\end{eqnarray}
where
$\rho=\sqrt{r^2+\alpha^2\cos^2{\theta}}$ and
$\Delta=r^2+\alpha^2-2Mr+q^2$.
The vierbein form can be read out from the line-element easily.
The spin form can be obtained by solving the torsionless condition with the above vierbein form:
\begin{eqnarray*}
\vomega_\mathrm{KN}^{\bullett}=\left(
\begin{array}{cccc}
0&\frac{r(\alpha^2+q^2-Mr)+(M-r)\alpha^2\cos^2{\theta}}{\sqrt{\Delta}\rho^3}\eee^0
+\frac{r\alpha\sin{\theta}}{\rho^3}\eee^3&
 \frac{\alpha^2\cos{\theta}\sin{\theta}}{\rho^3}\eee^0
 +\frac{\alpha\sqrt{\Delta}\cos{\theta}}{\rho^3}\eee^3&
\frac{\alpha r\sin{\theta}}{\rho^3}\eee^1
-\frac{\alpha\sqrt{\Delta}\cos{\theta}}{\rho^3}\eee^2\\
~& 0 & \frac{\alpha^2\cos{\theta}\sin{\theta}}{\rho^3}\eee^1
+\frac{\alpha\sqrt{\Delta}\cos{\theta}}{\rho^3}\eee^2 &
 \frac{\alpha r\sin{\theta}}{\rho^3}\eee^0
 +\frac{r\sqrt{\Delta}\cos{\theta}}{\rho^3}\eee^3\\
~&~&0&
\frac{\alpha\sqrt{\Delta}\sin{\theta}\cot{\theta}}{\rho^3}\eee^0
+\frac{(r^2+\alpha^2)\cot{\theta}}{\rho^3}\eee^3
\\
~&~&~&0\\
\end{array}
\right).
\end{eqnarray*}
The event horizon is given as one of the solutions of the equation $\Delta=0=(r-r_+)(r-r_-)$, which are given as
$
r_{\pm}=M\pm\sqrt{M^2-\alpha^2-q^2}.
$
The horizon corresponds to the larger solution $r_+$. 
A Kerr--Newman black hole has two Killing vectors of
$\xi_t=\partial_t,$ and $\xi_\phi=\partial_\phi$.
The Killing vector for the thermodynamic analysis is chosen to be $\xi=\beta_t\xi_t+\beta_\phi\xi_\phi$, where $\beta_t$ and  $\beta_\phi$ are the appropriate normalization factors.
We note that a linear combination of Killing vectors is also a Killing vector.   
Two cases of Killing vectors are considered in this work:
\begin{eqnarray}
\{\beta^t,\beta^\phi\}&=&
\left\{
\begin{array}{l}
\{1,\Omega_\mathrm{H}\},\\
\{0,c\},\label{kill}
\end{array}
\right.
\end{eqnarray}
where $\Omega_\mathrm{H}=\alpha/(r_+^2+\alpha^2)$ is the surface angular velocity.
A physical constant $c$ (the speed of light) is written explicitly in (\ref{kill}) for future convenience.
In the other parts, the physical constants are still set to unity.
Details of the treatment of the physical constants are given later in this section.
Even though each case corresponds to different thermodynamic processes, the calculations of the Noether charge are the same. 
Therefore, in the following, the $\beta$'s are not immediately specified.

The Noether charge with respect to this Killing vector has two non-zero surface forms of
\begin{eqnarray}
Q_\xi&=&\frac{1}{8\pi}
\int\QQQ_\xi
=\frac{1}{16\pi}\int
\left(
(\beta^a\omega^{r\phi}_a)\SSS_{r\phi}+
(\beta^a\omega^{tr}_a)\SSS_{tr}
\right),\label{nci}
\end{eqnarray}
where the suffix ``$a$'' can be $t$ or $\phi$.
Compared with the case of a Schwarzschild black hole, an additional factor of $1/2$ is placed in front of  $(\iota_\xi\vomega^{\bullett})$ to avoid double counting owing to the two surface forms that contribute to the results.
The other surface forms are zero at $\theta=\pi/2$ and therefore do not contribute to the final results.
These two terms have the following integrands:
\begin{eqnarray*}
\frac{1}{16\pi}
(\beta^a\omega^{r\phi}_a)\SSS_{r\phi}
=\frac{1}{16\pi}f_{t}(\theta;\beta^t,\beta^\phi)d\theta\wedge dt,&~&~~
\frac{1}{16\pi}
(\beta^a\omega^{tr}_a)\SSS_{tr}
=\frac{1}{16\pi}f_{\phi}(\theta;\beta^t,\beta^\phi)d\theta\wedge d\phi,
\end{eqnarray*}
where the integrand $f_\bullet(\theta;\beta^t,\beta^\phi)$ is given as;
\begin{eqnarray*}
f_{t,\phi}(x=\cos{\theta};\beta^t,\beta^\phi)&=&\frac{f_{t,\phi}^0
+f_{t,\phi}^2x^2+f_{t,\phi}^4x^4}{(r_+^2+\alpha^2x^2)^2}
\end{eqnarray*}
where
\begin{eqnarray*}
f_\phi^0&=&-r_+ \left(\alpha^2+r_+^2\right) 
\Bigl[
\alpha \beta^\phi r_+ (M+r_+)-\beta^t M r_++q^2 (\beta^t-\alpha  \beta^\phi)
\Bigr],\\
f_\phi^2&=&\alpha  \left(\alpha^2+r_+^2\right) 
\Bigl[
\beta^\phi \Bigl(r_+^2 (M+r_+)+\alpha^2 (M-r_+)-q^2 r_+\Bigr)-\alpha\beta^tM\Bigr],\\
f_\phi^4&=&-\alpha ^3 \beta^\phi (M-r_+) \left(\alpha^2+r^2\right),
\end{eqnarray*}
and
\begin{eqnarray*}
f_t^0&=&
\alpha  r_+ \left(\alpha \beta^\phi r_+ (M+r_+)-M \beta^t r_++q^2 (\beta^t-\alpha  )\right),\\
f_t^2&=&-\alpha^2 
\Bigl[
(\beta^\phi\left(r_+^2 (M+r_+)+\alpha^2 (M-r_+)-q^2 r_+\right)-\alpha\beta^t M\Bigr],\\
f_t^4&=&\alpha ^4 \beta^\phi (M-r_+).
\end{eqnarray*}
Following the $\theta$ and $\phi$ integrations, the Noether charge is obtained as
\begin{eqnarray}
Q_\xi&=&\frac{1}{16\pi}
\int\overline{Q}_{t}(\beta^t,\beta^\phi)dt+
\frac{1}{8}\overline{Q}_\phi(\beta^t,\beta^\phi),\label{QKN}
\end{eqnarray}
where $\overline{Q}_{t,\phi}(\beta^t,\beta^\phi)$ can be obtained by integrating a function $f_{t,\phi}(\theta;\beta^t,\beta^\phi)$.
The area of the event horizon of a Kerr--Newman black hole can be obtained as
\begin{eqnarray}
 A_\mathrm{KN}&=&\int\SSS_{tr}
=\left(r_+^2+\alpha^2\right)
 \int \sin{\theta}d\theta\wedge d\phi~=~4\pi\left(r_+^2+\alpha^2\right).\label{AKN}
\end{eqnarray}
Now, we can discuss the thermodynamics of a Kerr--Newman black hole.
\subsubsection{A Kerr black hole}
First, a Kerr black hole without an electric charge is treated, which can be obtained by setting $q\rightarrow0$. 
For the first case of (\ref{kill}), the Killing vector is null at the black hole surface
(\lq\lq{}event horizon\rq\rq{} = \rq\rq{}Killing horizon\rq\rq{}).
When $\alpha$ is zero, which corresponds to non-rotating black holes, the normalization is the same as that for a Schwarzschild black hole.
Therefore, in this case, the Noether charge is expected to be $T_\mathrm{KN}S_\mathrm{KN}$, where $T_\mathrm{KN}$ and $S_\mathrm{KN}$ are the temperature and the entropy of a Kerr--Newman black hole, respectively

The Noether charge in this case can be obtained as
\begin{eqnarray*}
\frac{c^2}{16\pi G}\overline{Q}_t(1,\Omega_\mathrm{K})\Bigl|_{q\rightarrow 0}=
-\frac{\pi c^3\Omega_\mathrm{K}}{2G} \sqrt{M'^2-\alpha ^2}dt,&~&~~
\frac{c^2}{16\pi G}\overline{Q}_\phi(1,\Omega_\mathrm{K})\Bigl|_{q\rightarrow 0}=
-\frac{c^2}{4G} \sqrt{M'^2-\alpha ^2},
\end{eqnarray*}
where $c\Omega_\mathrm{K}=\alpha c/(r_{K+}^2+\alpha^2)$ is the angular velocity of the event horizon and $r_{K+}=M'+\sqrt{M'^2-\alpha^2}$ is the diameter of the event horizon.
The definition and meaning of $M'$ is given below.
Here, physical parameters are put back into the expressions, which can be performed using the following replacements: 
$M\rightarrow M'=\frac{GM}{c^2}$,  $t~\rightarrow~ct$, $\xi_t~\rightarrow~\frac{1}{c}\xi_t$, 
and by multiplying a factor of $c^4/(4\pi G)$.
In this unit, $\alpha$ has ``length'' dimensions.
Therefore, one can confirm that the dimension of $c\Omega_\mathrm{K}$ is $t^{-1}$, which is consistent with its interpretation.
Because the angular velocity is constant on the event horizon (which is the zeroth law of black hole thermodynamics), this can be expressed as $d(\omega_\mathrm{K})/dt=c\Omega_\mathrm{K}$, where $\omega_\mathrm{K}$ is the rotation angle (which is constant) per unit time.
When an integration with respect to time is performed during one cycle of a black hole's rotation, one obtains 
\begin{eqnarray*}
\int\Omega_\mathrm{K}c~dt&=&\int\frac{d\omega_\mathrm{K}}{dt}dt~=
\int_0^{2\pi}d\omega_\mathrm{K}~=~2\pi.
\end{eqnarray*}
Therefore, the Noether charge can be obtained by summing two charges as
$
Q_\xi=-\frac{c^4}{2G} \sqrt{M'^2-\alpha ^2}=-T_\mathrm{K}S_\mathrm{K}
$.
Again, the temperature of the Kerr--Newman black hole,
\begin{eqnarray}
T_\mathrm{KN}&=&\frac{\hbar}{c} 
\frac{{r^2_{+}}-\alpha^2-q^2}{4\pi r_+(r^2_{+}+\alpha^2)}\label{tmpBH}
\end{eqnarray}
is borrowed from \cite{PhysRevD.15.2738} and is set to 
$T_\mathrm{K}=T_\mathrm{KN}(q\rightarrow 0)$ and $r_+\rightarrow r_{K+}$.
As a result, the entropy of the Kerr black hole can be extracted as
$
S_\mathrm{K}={A_\mathrm{K}}/{(4l_p^2)},
$
where
$
A_\mathrm{K}=8\pi M'\left(M'+\sqrt{M'^2-\alpha^2}\right).
$
Therefore, the area theorem for the black hole entropy is valid for a Kerr black hole. 

Next, let us take the second Killing vector, $\xi=c\partial_\phi$.
In this case, one can get
\begin{eqnarray*}
\frac{c^4}{16\pi G}\overline{Q}_t(0,c)\Bigl|_{q\rightarrow 0}=
\frac{1}{2}J_\mathrm{K}\Omega_\mathrm{K}c\frac{dt}{2\pi},&~&~~
\frac{c^4}{16\pi G}\overline{Q}_\phi(0,c)\Bigl|_{q\rightarrow 0}=\frac{1}{2}J_\mathrm{K},
\end{eqnarray*}
where $J_\mathrm{K}=Mc\alpha$ is the angular momentum of a black hole. 
After the $t$-integration, as before, the Noether charge becomes
$
Q_\xi=J_\mathrm{K}.
$
The Noether charge for this Killing vector consists of $J_\mathrm{K}$ and $\Omega_\mathrm{K}$, both of which are defined in classical physics, in contrast to the temperature, which can only be defined via the quantum effect.
Because the angular momentum is correctly obtained as the Noether charge, the method proposed here is ensured to work correctly.
Compared with the thermodynamics developed in Section \ref{sec5}, this may correspond to the isobaric process, which has a ``Killing vector'' of $\upsilon^a=(0,1,0)$.
The Noether charge induced by this Killing vector becomes $Q_\mrmTD=pV$.
This result supports the correspondence 
\begin{eqnarray}
pdV~\mathrm{(thermodynamics)}~\Leftrightarrow&~J_\mathrm{K}d\omega_\mathrm{K}~\mathrm{(black hole)}.
\end{eqnarray}

\subsubsection{A Kerr--Newman black hole} 
From (\ref{QKN}), the Noether charge for a Kerr--Newman black hole can be obtained as
\begin{eqnarray}
S_\mathrm{KN} T_\mathrm{KN}&=&-Q_\xi(1,\Omega_\mathrm{KN})
=\frac{c^4\sqrt{M^2-\alpha^2-q^2}}{2G},\label{KNST}
\end{eqnarray}
where the physical constants are written explicitly.
The charge $q$ is defined to have a length dimension.  
The entropy of a Kerr--Newman black hole can be singled out using the Hawking temperature (\ref{tmpBH}) as
\begin{eqnarray}
S_\mathrm{KN}&=&\frac{\pi c^3 r_+(r_+^2+\alpha^2)\sqrt{M'^2-\alpha^2-q^2}}{2(r_+^2-\alpha^2-q^2)G\hbar},\nonumber\\
&=&\frac{\pi c^3}{G\hbar}\left(
2M'(M'+\sqrt{M'^2-\alpha^2-q^2})-q^2
\right)=\frac{A_\mathrm{KN}}{4l_p^2}.\label{skn}
\end{eqnarray}
This corresponds to the area theorem for a Kerr--Newman black hole.

\subsection{Application to de Sitter space}
The de Sitter space is one of the cosmological solutions of the Einstein equation, which has a positive cosmological constant ($\Lambda>0$) without any matter or gauge fields.
The line element and vierbein forms of the static coordinate solution are similar to those of the Schwarzschild metrics, (\ref{lmschw}) and (\ref{virschw}). The result can be obtained from a simple replacement of
$
f^2_\mathrm{Schw}(r)\rightarrow f^2_\mathrm{dS}(r)=1-{r^2}/{L^2_\mathrm{dS}},
$
with the coordinate $x^a=(t,r,\theta,\phi)$, where $L_\mathrm{dS}=\sqrt{3/\Lambda}$ is the radius of the de Sitter horizon.
The physical constants are again written explicitly.
We note that the cosmological constant has a dimension of $(length$ $dimension)^{-2}$ in this convention.
According to this change, the spin form can be obtained as
\begin{eqnarray*}
\vomega_\mathrm{dS}^{\bullett}&=&
\left(
\begin{array}{cccc}
0&\r/L_\mathrm{dS}^{~2}~cdt~& 0 & 0 \\
~& 0 &  f_\mathrm{dS}d\theta &
f_\mathrm{dS}\sin{\theta}~d\phi \\
~&~&0&\cos{\theta}~d\phi\\
~&~&~&0\\
\end{array}
\right).
\end{eqnarray*}
Therefore, the Noether charge with respect to the Killing vector $\xi_t=\partial_t/c$ can be calculated to be
\begin{eqnarray*}
\frac{c^4}{8\pi G}\int\QQQ_{\xi_t}&=&\frac{c^4}{16\pi G}\int2L_\mathrm{dS}\sin{\theta}~d\theta\wedge d\phi
=\frac{c^4L_\mathrm{dS}}{2G}.
\end{eqnarray*}
Together with the surface temperature of the horizon, such that
\begin{eqnarray*}
T_\mathrm{dS}&=&\frac{\hbar}{c}\frac{\kappa_\mathrm{dS}}{2\pi}~=~\frac{\hbar}{2\pi c}\frac{1}{L_\mathrm{dS}},
\end{eqnarray*}
the entropy can be obtained as
\begin{eqnarray*}
S_\mathrm{dS}&=&\frac{c^4L_\mathrm{dS}}{2GT_\mathrm{dS}}~=~\frac{A_\mathrm{dS}}{4l^2_p},
\end{eqnarray*}
where $A_\mathrm{dS}=4\pi L^2_\mathrm{dS}$ is the surface area of the de Sitter horizon.
The area theorem is also valid in this case.
\section{Concluding remarks}\label{sec7}
The method proposed in this work, called the ``{\it geometrothermodynamic method} (GT-method)'', is sufficiently general to treat a large variety of gravitational objects that have event horizons, such as black holes and de Sitter space. 
The Noether charge associated with an appropriate Killing vector shows a clear relationship with thermodynamical objects, such as the temperature and entropy.
One advantage of the GT-method is that it is a purely classical and thermodynamic method that does not assume any (Euclidean) ensembles.
Moreover, its formalism is highly geometrical and can be described in a coordinate independent way.
The same formalism can also be applied to thermodynamics, and therefore the relationships between the thermodynamics and gravitational theory are apparent.
The resultant entropies obtained using the GT-method are consistent with previous results and confirm the area theorem for black holes and the de Sitter space.

To understand the GT-method more fully, let us recall a method proposed in \cite{Bardeen} and discussed in \cite{0264-9381-17-2-310}.
The area of the event horizon of a Kerr--Newman black hole is again shown here as
$
 A=4\pi\left(
 2M\left(M+\sqrt{M^2-(J/M)^2-q^2}\right)-q^2
 \right)
$,
where the subscript ``KN'' is omitted from the expression.
Here, this relation is considered to be a state equation with three independent variables, $M$, $J$, and $q$, and has a variation with respect to these variables:
\begin{eqnarray*}
{\delta A(M,J,q)}&=& \frac{\partial A}{\partial M}{\delta M}
+ \frac{\partial A}{\partial J}\delta J
+ \frac{\partial A}{\partial q}{\delta q}.
\end{eqnarray*}
One can see after simple manipulations that
\begin{eqnarray}
-{\delta M}&=&-F \delta A+\Omega dJ+\Phi dq,\label{stateeq}
\end{eqnarray}
where
\begin{eqnarray*}
F=\frac{\sqrt{M^2-\alpha^2-q^2}}{2A},~~
\Omega=\frac{J/M}{r_+^2+(J/M)^2},~~
\Phi=\frac{qr_+}{r_+^2+(J/M)^2}.
\end{eqnarray*}
Here $\Omega$ and $\Phi$ correctly correspond to the surface velocity and the electric potential, respectively.  
A question that needs to be asked here is what is the relationship between $F$ and the temperature/entropy.
For any black hole, the Hawking temperature and the area of the horizon have the relationship of
$
TA={(r_+^2-\alpha^2-q^2)}/{r_+}.
$
Therefore, one can easily confirm the relation
$
F={T}/{4}
$.
Therefore, the first term of the r.h.s. of (\ref{stateeq}) can be written as
\begin{eqnarray*}
-F \delta A&=&-T\frac{\delta A}{4}~=-T{\delta S},
\end{eqnarray*}
with $S=A/4$.
This simple analysis also verifies the area theorem and the first law of black hole thermodynamics.
The implicit assumption in this discussion is that the term $F\delta A$ corresponds to the $S\delta T$ term in thermodynamics.
If we accept that the entropy of a black hole is only a function of the area and that the temperature is given by the Hawking radiation, the differential equation
$
{dS(A)}/{dA}={1}/{4}
$,
with a boundary condition of $S=0$ at $A=0$ uniquely gives the solution $S=A/4$.

Conversely, the GT-method does not assume any relationship between the entropy and the area.
The area of the event horizon appears in non-zero components of the surface form $\SSS_{\bullett}$ in the Noether charge.
A second term on the r.h.s$.$ of (\ref{stateeq}) is given by the same procedure with a different Killing vector.
The reason why the Noether charge is proportional to the surface form is very simple.
In diffeomorphic theory for a $d$-dimensional manifold, the Lagrangian form and the Noether current must be the $d$-form and the $(d-1)$-form, respectively. 
Therefore, the Noether charge, which is obtained by integrating the Noether current,  must be a $(d-2)$-form.
In four-dimensional space-time, the Noether charge is a two-form, which can be expanded by the bases $\eee^{a_1}\wedge\eee^{a_2}=\overline{\SSS}^{a_1a_2}$.
Therefore, the GT-method provides independent evidence that the entropy can be represented using only the area.

In summary, the GT-method, which can extract thermodynamic quantities from a large variety of solutions for the Einstein equation with an event horizon, is proposed in this work.
This method is mathematically consistent with Wald's method.
However, the relationships between general relativity and thermodynamics are clearer because abstract thermodynamics can be reconstructed using geometrical terminologies that are parallel to general relativity.


The discussions given in this work are purely thermodynamic without any statistical (Euclidean) ensembles.
Therefore, the conclusion, for example, that $S_\mathrm{KN}T_\mathrm{KN}$ of a Kerr--Newman black hole, such as (\ref{KNST}), must be independent of the microscopic details is expected to be correct, even if the expression of the Hawking temperature will receive some corrections from the (still-unknown) quantum gravity.  
\vskip 0.5cm
We wish to thank Dr. Y.~Sugiyama for their continuous encouragement and fruitful discussions.
\appendix
\section{Examples of the Noether charge}\label{app1}
\subsection{Kruscal--Szekeres coordinate}
A Shwarzschild black hole can be expressed using Kruscal--Szekers coordinates\footnote{See, for example, section $2.2$ in ref.\cite{fre2012gravity2}.} instead of a standard polar coordinate system. 
A line element squared can be written in Kruscal--Szekers coordinates $x^\mu=(T, X, \theta, \phi)$ as
\begin{eqnarray}
ds^2&=&4\frac{r_s^3}{r}\exp{\left(-\frac{r}{r_s}\right)}\left(dT^2-dX^2\right)-r^2d\theta^2-r^2\sin^2{\theta}d\phi^2,
\end{eqnarray}
where $r_s=2GM$ is a radius of an event horizon.
A relation between four dimensional Kruscal--Szekers and polar coordinates is given as;
\begin{eqnarray}
\left\{
\begin{array}{l}
X^2-T^2=\left(\frac{r}{r_s}-1\right)\exp{\left(\frac{r}{r_s}\right)},\\
2\tanh{\left(\frac{T}{X}\right)}=\frac{t}{r_s},\\
\theta=\theta,\\
\phi=\phi.
\end{array}
\right.\label{KS}
\end{eqnarray}
The Noether charge in general relativity (\ref{GRNC}) can be expressed in this coordinates for the vector field $\xi^\mu=(\xi_T,\xi_X,\xi_\theta,\xi_\phi)$ as;
\begin{eqnarray}
\left\{
\begin{array}{lcc}
\xi_T&:&-4\pi\exp{\left(-\frac{r}{r_s}\right)}r_s\left(r+r_s\right)X,\\
\xi_X&:&4\pi\exp{\left(-\frac{r}{r_s}\right)}r_s\left(r+r_s\right)T,\\
\xi_\theta&:&4\pi\exp{\left(-\frac{r}{r_s}\right)}r_s^2\left(TdX-XdT\right),\\
\xi_\phi&:&-4\exp{\left(-\frac{r}{r_s}\right)}r_s^2\left(TdX-XdT\right),
\end{array}
\right.
\end{eqnarray}
where $\theta$ and $\phi$ are integrated out, and a bare $\theta$  (not in a $\theta$ integration) is set to $\pi/2$. 
The function $r=r(T,X)$ can be obtained as a solution of (\ref{KS}).

\subsection{Friedmann-Lme\^itre-Robertson-Walker metric}
The Friedmann-Lme\^itre-Robertson-Walker (FLRW) metric\cite{Einstein1922,Lmeitre,1935ApJ82284R,Walker01011937} is a homogeneous and isotropic solution of the Einstein equation whose line element squared is given as;
\begin{eqnarray}
ds^2&=&dt^2-\Omega(t)^2\left(f^{-1}(r)dr+r^2d\theta^2+r^2\sin^2{\theta}d\phi^2\right),
\end{eqnarray}
where $f(r)=1-K r^2$ and $\Omega(t)$ is a scale function.
The Noether charge can be obtained in this coordinates for the vector field $\xi^\mu=(\xi_t,\xi_r,\xi_\theta,\xi_\phi)$ as;
\begin{eqnarray}
\left\{
\begin{array}{lcc}
\xi_t&:&0,\\
\xi_r&:&4\pi r^2 f(r)^{-\frac{1}{2}}\Omega(t)^2{\dot\Omega(t)},\\
\xi_\theta&:&-2\pi rf(r)^{\frac{1}{2}}\Omega(t)dt-2\pi r^2f(r)^{-\frac{1}{2}}\Omega(t)^2{\dot\Omega(t)},\\
\xi_\phi&:&2rf(r)^{\frac{1}{2}}\Omega(t)dt+2r^2f(r)^{-\frac{1}{2}}\Omega(t)^2{\dot\Omega(t)}dr,
\end{array}
\right.
\end{eqnarray}
where ${\dot\Omega(t)}=d\Omega(t)/dt$.
Here $\theta$ and $\phi$ are integrated out, and a bare $\theta$  is set to $\pi/2$.

%
\bibliography{ref}

\begin{thebibliography}{69}%
\makeatletter
\providecommand \@ifxundefined [1]{%
 \@ifx{#1\undefined}
}%
\providecommand \@ifnum [1]{%
 \ifnum #1\expandafter \@firstoftwo
 \else \expandafter \@secondoftwo
 \fi
}%
\providecommand \@ifx [1]{%
 \ifx #1\expandafter \@firstoftwo
 \else \expandafter \@secondoftwo
 \fi
}%
\providecommand \natexlab [1]{#1}%
\providecommand \enquote  [1]{``#1''}%
\providecommand \bibnamefont  [1]{#1}%
\providecommand \bibfnamefont [1]{#1}%
\providecommand \citenamefont [1]{#1}%
\providecommand \href@noop [0]{\@secondoftwo}%
\providecommand \href [0]{\begingroup \@sanitize@url \@href}%
\providecommand \@href[1]{\@@startlink{#1}\@@href}%
\providecommand \@@href[1]{\endgroup#1\@@endlink}%
\providecommand \@sanitize@url [0]{\catcode `\\12\catcode `\$12\catcode
  `\&12\catcode `\#12\catcode `\^12\catcode `\_12\catcode `\%12\relax}%
\providecommand \@@startlink[1]{}%
\providecommand \@@endlink[0]{}%
\providecommand \url  [0]{\begingroup\@sanitize@url \@url }%
\providecommand \@url [1]{\endgroup\@href {#1}{\urlprefix }}%
\providecommand \urlprefix  [0]{URL }%
\providecommand \Eprint [0]{\href }%
\providecommand \doibase [0]{http://dx.doi.org/}%
\providecommand \selectlanguage [0]{\@gobble}%
\providecommand \bibinfo  [0]{\@secondoftwo}%
\providecommand \bibfield  [0]{\@secondoftwo}%
\providecommand \translation [1]{[#1]}%
\providecommand \BibitemOpen [0]{}%
\providecommand \bibitemStop [0]{}%
\providecommand \bibitemNoStop [0]{.\EOS\space}%
\providecommand \EOS [0]{\spacefactor3000\relax}%
\providecommand \BibitemShut  [1]{\csname bibitem#1\endcsname}%
\let\auto@bib@innerbib\@empty
\bibitem [{\citenamefont {Bekenstein}(1973)}]{PhysRevD.7.2333}%
  \BibitemOpen
  \bibfield  {author} {\bibinfo {author} {\bibfnamefont {J.~D.}\ \bibnamefont
  {Bekenstein}},\ }\href {\doibase 10.1103/PhysRevD.7.2333} {\bibfield
  {journal} {\bibinfo  {journal} {Phys. Rev. D}\ }\textbf {\bibinfo {volume}
  {7}},\ \bibinfo {pages} {2333} (\bibinfo {year} {1973})}\BibitemShut
  {NoStop}%
\bibitem [{\citenamefont {Hawking}(1975)}]{hawking1975}%
  \BibitemOpen
  \bibfield  {author} {\bibinfo {author} {\bibfnamefont {S.~W.}\ \bibnamefont
  {Hawking}},\ }\href {\doibase 10.1007/BF02345020} {\bibfield  {journal}
  {\bibinfo  {journal} {Communications in Mathematical Physics}\ }\textbf
  {\bibinfo {volume} {43}},\ \bibinfo {pages} {199} (\bibinfo {year}
  {1975})}\BibitemShut {NoStop}%
\bibitem [{\citenamefont {Bardeen}\ \emph {et~al.}()\citenamefont {Bardeen},
  \citenamefont {Carter},\ and\ \citenamefont {Hawking}}]{Bardeen}%
  \BibitemOpen
  \bibfield  {author} {\bibinfo {author} {\bibfnamefont {J.~M.}\ \bibnamefont
  {Bardeen}}, \bibinfo {author} {\bibfnamefont {B.}~\bibnamefont {Carter}}, \
  and\ \bibinfo {author} {\bibfnamefont {S.~W.}\ \bibnamefont {Hawking}},\
  }\href {\doibase 10.1007/BF01645742} {\bibfield  {journal} {\bibinfo
  {journal} {Communications in Mathematical Physics}\ }\textbf {\bibinfo
  {volume} {31}},\ \bibinfo {pages} {161}}\BibitemShut {NoStop}%
\bibitem [{\citenamefont {Carter}(1971)}]{PhysRevLett.26.331}%
  \BibitemOpen
  \bibfield  {author} {\bibinfo {author} {\bibfnamefont {B.}~\bibnamefont
  {Carter}},\ }\href {\doibase 10.1103/PhysRevLett.26.331} {\bibfield
  {journal} {\bibinfo  {journal} {Phys. Rev. Lett.}\ }\textbf {\bibinfo
  {volume} {26}},\ \bibinfo {pages} {331} (\bibinfo {year} {1971})}\BibitemShut
  {NoStop}%
\bibitem [{\citenamefont {Wald}(1995)}]{WALD9507055}%
  \BibitemOpen
  \bibfield  {author} {\bibinfo {author} {\bibfnamefont {R.}~\bibnamefont
  {Wald}},\ }\href@noop {} {\bibfield  {journal} {\bibinfo  {journal}
  {gr-qc/9507055}\ } (\bibinfo {year} {1995})}\BibitemShut {NoStop}%
\bibitem [{\citenamefont {Gould}(1987)}]{PhysRevD.35.449}%
  \BibitemOpen
  \bibfield  {author} {\bibinfo {author} {\bibfnamefont {A.}~\bibnamefont
  {Gould}},\ }\href {\doibase 10.1103/PhysRevD.35.449} {\bibfield  {journal}
  {\bibinfo  {journal} {Phys. Rev. D}\ }\textbf {\bibinfo {volume} {35}},\
  \bibinfo {pages} {449} (\bibinfo {year} {1987})}\BibitemShut {NoStop}%
\bibitem [{\citenamefont {Wald}(1993)}]{PhysRevD.48.R3427}%
  \BibitemOpen
  \bibfield  {author} {\bibinfo {author} {\bibfnamefont {R.~M.}\ \bibnamefont
  {Wald}},\ }\href {\doibase 10.1103/PhysRevD.48.R3427} {\bibfield  {journal}
  {\bibinfo  {journal} {Phys. Rev. D}\ }\textbf {\bibinfo {volume} {48}},\
  \bibinfo {pages} {R3427} (\bibinfo {year} {1993})}\BibitemShut {NoStop}%
\bibitem [{\citenamefont {Iyer}\ and\ \citenamefont
  {Wald}(1994)}]{PhysRevD.50.846}%
  \BibitemOpen
  \bibfield  {author} {\bibinfo {author} {\bibfnamefont {V.}~\bibnamefont
  {Iyer}}\ and\ \bibinfo {author} {\bibfnamefont {R.~M.}\ \bibnamefont
  {Wald}},\ }\href {\doibase 10.1103/PhysRevD.50.846} {\bibfield  {journal}
  {\bibinfo  {journal} {Phys. Rev. D}\ }\textbf {\bibinfo {volume} {50}},\
  \bibinfo {pages} {846} (\bibinfo {year} {1994})}\BibitemShut {NoStop}%
\bibitem [{\citenamefont {Iyer}\ and\ \citenamefont
  {Wald}(1995)}]{PhysRevD.52.4430}%
  \BibitemOpen
  \bibfield  {author} {\bibinfo {author} {\bibfnamefont {V.}~\bibnamefont
  {Iyer}}\ and\ \bibinfo {author} {\bibfnamefont {R.~M.}\ \bibnamefont
  {Wald}},\ }\href {\doibase 10.1103/PhysRevD.52.4430} {\bibfield  {journal}
  {\bibinfo  {journal} {Phys. Rev. D}\ }\textbf {\bibinfo {volume} {52}},\
  \bibinfo {pages} {4430} (\bibinfo {year} {1995})}\BibitemShut {NoStop}%
\bibitem [{\citenamefont {Wald}\ and\ \citenamefont
  {Zoupas}(2000)}]{PhysRevD.61.084027}%
  \BibitemOpen
  \bibfield  {author} {\bibinfo {author} {\bibfnamefont {R.~M.}\ \bibnamefont
  {Wald}}\ and\ \bibinfo {author} {\bibfnamefont {A.}~\bibnamefont {Zoupas}},\
  }\href {\doibase 10.1103/PhysRevD.61.084027} {\bibfield  {journal} {\bibinfo
  {journal} {Phys. Rev. D}\ }\textbf {\bibinfo {volume} {61}},\ \bibinfo
  {pages} {084027} (\bibinfo {year} {2000})}\BibitemShut {NoStop}%
\bibitem [{\citenamefont {'t~Hooft}(1985)}]{THOOFT1985727}%
  \BibitemOpen
  \bibfield  {author} {\bibinfo {author} {\bibfnamefont {G.}~\bibnamefont
  {'t~Hooft}},\ }\href {\doibase https://doi.org/10.1016/0550-3213(85)90418-3}
  {\bibfield  {journal} {\bibinfo  {journal} {Nuclear Physics B}\ }\textbf
  {\bibinfo {volume} {256}},\ \bibinfo {pages} {727 } (\bibinfo {year}
  {1985})}\BibitemShut {NoStop}%
\bibitem [{\citenamefont {Bombelli}\ \emph {et~al.}(1986)\citenamefont
  {Bombelli}, \citenamefont {Koul}, \citenamefont {Lee},\ and\ \citenamefont
  {Sorkin}}]{PhysRevD.34.373}%
  \BibitemOpen
  \bibfield  {author} {\bibinfo {author} {\bibfnamefont {L.}~\bibnamefont
  {Bombelli}}, \bibinfo {author} {\bibfnamefont {R.~K.}\ \bibnamefont {Koul}},
  \bibinfo {author} {\bibfnamefont {J.}~\bibnamefont {Lee}}, \ and\ \bibinfo
  {author} {\bibfnamefont {R.~D.}\ \bibnamefont {Sorkin}},\ }\href {\doibase
  10.1103/PhysRevD.34.373} {\bibfield  {journal} {\bibinfo  {journal} {Phys.
  Rev. D}\ }\textbf {\bibinfo {volume} {34}},\ \bibinfo {pages} {373} (\bibinfo
  {year} {1986})}\BibitemShut {NoStop}%
\bibitem [{\citenamefont {Brown}\ and\ \citenamefont
  {York}(1993)}]{PhysRevD.47.1420}%
  \BibitemOpen
  \bibfield  {author} {\bibinfo {author} {\bibfnamefont {J.~D.}\ \bibnamefont
  {Brown}}\ and\ \bibinfo {author} {\bibfnamefont {J.~W.}\ \bibnamefont
  {York}},\ }\href {\doibase 10.1103/PhysRevD.47.1420} {\bibfield  {journal}
  {\bibinfo  {journal} {Phys. Rev. D}\ }\textbf {\bibinfo {volume} {47}},\
  \bibinfo {pages} {1420} (\bibinfo {year} {1993})}\BibitemShut {NoStop}%
\bibitem [{\citenamefont {Jacobson}\ \emph {et~al.}(1994)\citenamefont
  {Jacobson}, \citenamefont {Kang},\ and\ \citenamefont
  {Myers}}]{PhysRevD.49.6587}%
  \BibitemOpen
  \bibfield  {author} {\bibinfo {author} {\bibfnamefont {T.}~\bibnamefont
  {Jacobson}}, \bibinfo {author} {\bibfnamefont {G.}~\bibnamefont {Kang}}, \
  and\ \bibinfo {author} {\bibfnamefont {R.~C.}\ \bibnamefont {Myers}},\ }\href
  {\doibase 10.1103/PhysRevD.49.6587} {\bibfield  {journal} {\bibinfo
  {journal} {Phys. Rev. D}\ }\textbf {\bibinfo {volume} {49}},\ \bibinfo
  {pages} {6587} (\bibinfo {year} {1994})}\BibitemShut {NoStop}%
\bibitem [{\citenamefont {Callan}\ and\ \citenamefont
  {Wilczek}(1994)}]{CALLAN199455}%
  \BibitemOpen
  \bibfield  {author} {\bibinfo {author} {\bibfnamefont {C.}~\bibnamefont
  {Callan}}\ and\ \bibinfo {author} {\bibfnamefont {F.}~\bibnamefont
  {Wilczek}},\ }\href {\doibase https://doi.org/10.1016/0370-2693(94)91007-3}
  {\bibfield  {journal} {\bibinfo  {journal} {Physics Letters B}\ }\textbf
  {\bibinfo {volume} {333}},\ \bibinfo {pages} {55 } (\bibinfo {year}
  {1994})}\BibitemShut {NoStop}%
\bibitem [{\citenamefont {Holzhey}\ \emph {et~al.}(1994)\citenamefont
  {Holzhey}, \citenamefont {Larsen},\ and\ \citenamefont
  {Wilczek}}]{HOLZHEY1994443}%
  \BibitemOpen
  \bibfield  {author} {\bibinfo {author} {\bibfnamefont {C.}~\bibnamefont
  {Holzhey}}, \bibinfo {author} {\bibfnamefont {F.}~\bibnamefont {Larsen}}, \
  and\ \bibinfo {author} {\bibfnamefont {F.}~\bibnamefont {Wilczek}},\ }\href
  {\doibase https://doi.org/10.1016/0550-3213(94)90402-2} {\bibfield  {journal}
  {\bibinfo  {journal} {Nuclear Physics B}\ }\textbf {\bibinfo {volume}
  {424}},\ \bibinfo {pages} {443 } (\bibinfo {year} {1994})}\BibitemShut
  {NoStop}%
\bibitem [{\citenamefont {Liberati}(1997)}]{Liberati:1996kt}%
  \BibitemOpen
  \bibfield  {author} {\bibinfo {author} {\bibfnamefont {S.}~\bibnamefont
  {Liberati}},\ }\bibfield  {booktitle} {\emph {\bibinfo {booktitle}
  {{Relativistic astrophysics. Proceedings, 4th Italian-Korean Meeting, Rome,
  Gran Sasso, Pescara, Italy, July 9-15, 1995}}},\ }\href@noop {} {\bibfield
  {journal} {\bibinfo  {journal} {Nuovo Cim.}\ }\textbf {\bibinfo {volume}
  {B112}},\ \bibinfo {pages} {405} (\bibinfo {year} {1997})},\ \Eprint
  {http://arxiv.org/abs/gr-qc/9601032} {arXiv:gr-qc/9601032 [gr-qc]}
  \BibitemShut {NoStop}%
\bibitem [{\citenamefont {Ashtekar}\ \emph {et~al.}(1998)\citenamefont
  {Ashtekar}, \citenamefont {Baez}, \citenamefont {Corichi},\ and\
  \citenamefont {Krasnov}}]{PhysRevLett.80.904}%
  \BibitemOpen
  \bibfield  {author} {\bibinfo {author} {\bibfnamefont {A.}~\bibnamefont
  {Ashtekar}}, \bibinfo {author} {\bibfnamefont {J.}~\bibnamefont {Baez}},
  \bibinfo {author} {\bibfnamefont {A.}~\bibnamefont {Corichi}}, \ and\
  \bibinfo {author} {\bibfnamefont {K.}~\bibnamefont {Krasnov}},\ }\href
  {\doibase 10.1103/PhysRevLett.80.904} {\bibfield  {journal} {\bibinfo
  {journal} {Phys. Rev. Lett.}\ }\textbf {\bibinfo {volume} {80}},\ \bibinfo
  {pages} {904} (\bibinfo {year} {1998})}\BibitemShut {NoStop}%
\bibitem [{\citenamefont {Carlip}(2007)}]{Carlip:2007qh}%
  \BibitemOpen
  \bibfield  {author} {\bibinfo {author} {\bibfnamefont {S.}~\bibnamefont
  {Carlip}},\ }\href {\doibase 10.1142/S0218271808012401,
  10.1007/s10714-007-0467-6} {\bibfield  {journal} {\bibinfo  {journal} {Gen.
  Rel. Grav.}\ }\textbf {\bibinfo {volume} {39}},\ \bibinfo {pages} {1519}
  (\bibinfo {year} {2007})},\ \bibinfo {note} {[Int. J. Mod.
  Phys.D17,659(2008)]},\ \Eprint {http://arxiv.org/abs/0705.3024}
  {arXiv:0705.3024 [gr-qc]} \BibitemShut {NoStop}%
\bibitem [{\citenamefont {Carroll}\ \emph {et~al.}(2009)\citenamefont
  {Carroll}, \citenamefont {Johnson},\ and\ \citenamefont
  {Randall}}]{1126-6708-2009-11-109}%
  \BibitemOpen
  \bibfield  {author} {\bibinfo {author} {\bibfnamefont {S.~M.}\ \bibnamefont
  {Carroll}}, \bibinfo {author} {\bibfnamefont {M.~C.}\ \bibnamefont
  {Johnson}}, \ and\ \bibinfo {author} {\bibfnamefont {L.}~\bibnamefont
  {Randall}},\ }\href {http://stacks.iop.org/1126-6708/2009/i=11/a=109}
  {\bibfield  {journal} {\bibinfo  {journal} {Journal of High Energy Physics}\
  }\textbf {\bibinfo {volume} {2009}},\ \bibinfo {pages} {109} (\bibinfo {year}
  {2009})}\BibitemShut {NoStop}%
\bibitem [{\citenamefont {Kolekar}\ \emph {et~al.}(2012)\citenamefont
  {Kolekar}, \citenamefont {Kothawala},\ and\ \citenamefont
  {Padmanabhan}}]{PhysRevD.85.064031}%
  \BibitemOpen
  \bibfield  {author} {\bibinfo {author} {\bibfnamefont {S.}~\bibnamefont
  {Kolekar}}, \bibinfo {author} {\bibfnamefont {D.}~\bibnamefont {Kothawala}},
  \ and\ \bibinfo {author} {\bibfnamefont {T.}~\bibnamefont {Padmanabhan}},\
  }\href {\doibase 10.1103/PhysRevD.85.064031} {\bibfield  {journal} {\bibinfo
  {journal} {Phys. Rev. D}\ }\textbf {\bibinfo {volume} {85}},\ \bibinfo
  {pages} {064031} (\bibinfo {year} {2012})}\BibitemShut {NoStop}%
\bibitem [{\citenamefont {Compère}\ \emph {et~al.}(2015)\citenamefont
  {Compère}, \citenamefont {Hajian}, \citenamefont {Seraj},\ and\
  \citenamefont {Sheikh-Jabbari}}]{COMPERE2015443}%
  \BibitemOpen
  \bibfield  {author} {\bibinfo {author} {\bibfnamefont {G.}~\bibnamefont
  {Compère}}, \bibinfo {author} {\bibfnamefont {K.}~\bibnamefont {Hajian}},
  \bibinfo {author} {\bibfnamefont {A.}~\bibnamefont {Seraj}}, \ and\ \bibinfo
  {author} {\bibfnamefont {M.}~\bibnamefont {Sheikh-Jabbari}},\ }\href
  {\doibase https://doi.org/10.1016/j.physletb.2015.08.027} {\bibfield
  {journal} {\bibinfo  {journal} {Physics Letters B}\ }\textbf {\bibinfo
  {volume} {749}},\ \bibinfo {pages} {443 } (\bibinfo {year}
  {2015})}\BibitemShut {NoStop}%
\bibitem [{\citenamefont {Comp{\`e}re}\ \emph {et~al.}(2015)\citenamefont
  {Comp{\`e}re}, \citenamefont {Hajian}, \citenamefont {Seraj},\ and\
  \citenamefont {Sheikh-Jabbari}}]{Compare2015}%
  \BibitemOpen
  \bibfield  {author} {\bibinfo {author} {\bibfnamefont {G.}~\bibnamefont
  {Comp{\`e}re}}, \bibinfo {author} {\bibfnamefont {K.}~\bibnamefont {Hajian}},
  \bibinfo {author} {\bibfnamefont {A.}~\bibnamefont {Seraj}}, \ and\ \bibinfo
  {author} {\bibfnamefont {M.~M.}\ \bibnamefont {Sheikh-Jabbari}},\ }\href
  {\doibase 10.1007/JHEP10(2015)093} {\bibfield  {journal} {\bibinfo  {journal}
  {Journal of High Energy Physics}\ }\textbf {\bibinfo {volume} {2015}},\
  \bibinfo {pages} {93} (\bibinfo {year} {2015})}\BibitemShut {NoStop}%
\bibitem [{\citenamefont {Saravani}\ \emph {et~al.}(2015)\citenamefont
  {Saravani}, \citenamefont {Afshordi},\ and\ \citenamefont
  {Mann}}]{Saravani:2012is}%
  \BibitemOpen
  \bibfield  {author} {\bibinfo {author} {\bibfnamefont {M.}~\bibnamefont
  {Saravani}}, \bibinfo {author} {\bibfnamefont {N.}~\bibnamefont {Afshordi}},
  \ and\ \bibinfo {author} {\bibfnamefont {R.~B.}\ \bibnamefont {Mann}},\
  }\href {\doibase 10.1142/S021827181443007X} {\bibfield  {journal} {\bibinfo
  {journal} {Int. J. Mod. Phys.}\ }\textbf {\bibinfo {volume} {D23}},\ \bibinfo
  {pages} {1443007} (\bibinfo {year} {2015})},\ \Eprint
  {http://arxiv.org/abs/1212.4176} {arXiv:1212.4176 [hep-th]} \BibitemShut
  {NoStop}%
\bibitem [{\citenamefont {Cai}\ \emph {et~al.}(2015)\citenamefont {Cai},
  \citenamefont {Hu}, \citenamefont {Pan},\ and\ \citenamefont
  {Zhang}}]{PhysRevD.91.024032}%
  \BibitemOpen
  \bibfield  {author} {\bibinfo {author} {\bibfnamefont {R.-G.}\ \bibnamefont
  {Cai}}, \bibinfo {author} {\bibfnamefont {Y.-P.}\ \bibnamefont {Hu}},
  \bibinfo {author} {\bibfnamefont {Q.-Y.}\ \bibnamefont {Pan}}, \ and\
  \bibinfo {author} {\bibfnamefont {Y.-L.}\ \bibnamefont {Zhang}},\ }\href
  {\doibase 10.1103/PhysRevD.91.024032} {\bibfield  {journal} {\bibinfo
  {journal} {Phys. Rev. D}\ }\textbf {\bibinfo {volume} {91}},\ \bibinfo
  {pages} {024032} (\bibinfo {year} {2015})}\BibitemShut {NoStop}%
\bibitem [{\citenamefont {Strominger}\ and\ \citenamefont
  {Vafa}(1996)}]{STROMINGER199699}%
  \BibitemOpen
  \bibfield  {author} {\bibinfo {author} {\bibfnamefont {A.}~\bibnamefont
  {Strominger}}\ and\ \bibinfo {author} {\bibfnamefont {C.}~\bibnamefont
  {Vafa}},\ }\href {\doibase https://doi.org/10.1016/0370-2693(96)00345-0}
  {\bibfield  {journal} {\bibinfo  {journal} {Physics Letters B}\ }\textbf
  {\bibinfo {volume} {379}},\ \bibinfo {pages} {99 } (\bibinfo {year}
  {1996})}\BibitemShut {NoStop}%
\bibitem [{\citenamefont {Lunin}\ and\ \citenamefont
  {Mathur}(2002)}]{LUNIN2002342}%
  \BibitemOpen
  \bibfield  {author} {\bibinfo {author} {\bibfnamefont {O.}~\bibnamefont
  {Lunin}}\ and\ \bibinfo {author} {\bibfnamefont {S.~D.}\ \bibnamefont
  {Mathur}},\ }\href {\doibase https://doi.org/10.1016/S0550-3213(01)00620-4}
  {\bibfield  {journal} {\bibinfo  {journal} {Nuclear Physics B}\ }\textbf
  {\bibinfo {volume} {623}},\ \bibinfo {pages} {342 } (\bibinfo {year}
  {2002})}\BibitemShut {NoStop}%
\bibitem [{\citenamefont {Ryu}\ and\ \citenamefont
  {Takayanagi}(2006)}]{PhysRevLett.96.181602}%
  \BibitemOpen
  \bibfield  {author} {\bibinfo {author} {\bibfnamefont {S.}~\bibnamefont
  {Ryu}}\ and\ \bibinfo {author} {\bibfnamefont {T.}~\bibnamefont
  {Takayanagi}},\ }\href {\doibase 10.1103/PhysRevLett.96.181602} {\bibfield
  {journal} {\bibinfo  {journal} {Phys. Rev. Lett.}\ }\textbf {\bibinfo
  {volume} {96}},\ \bibinfo {pages} {181602} (\bibinfo {year}
  {2006})}\BibitemShut {NoStop}%
\bibitem [{\citenamefont {Hubeny}\ \emph {et~al.}(2007)\citenamefont {Hubeny},
  \citenamefont {Rangamani},\ and\ \citenamefont {Takayanagi}}]{Hubeny:2007xt}%
  \BibitemOpen
  \bibfield  {author} {\bibinfo {author} {\bibfnamefont {V.~E.}\ \bibnamefont
  {Hubeny}}, \bibinfo {author} {\bibfnamefont {M.}~\bibnamefont {Rangamani}}, \
  and\ \bibinfo {author} {\bibfnamefont {T.}~\bibnamefont {Takayanagi}},\
  }\href {\doibase 10.1088/1126-6708/2007/07/062} {\bibfield  {journal}
  {\bibinfo  {journal} {JHEP}\ }\textbf {\bibinfo {volume} {07}},\ \bibinfo
  {pages} {062} (\bibinfo {year} {2007})},\ \Eprint
  {http://arxiv.org/abs/0705.0016} {arXiv:0705.0016 [hep-th]} \BibitemShut
  {NoStop}%
\bibitem [{\citenamefont {Comp{\`e}re}\ \emph {et~al.}(2016)\citenamefont
  {Comp{\`e}re}, \citenamefont {Mao}, \citenamefont {Seraj},\ and\
  \citenamefont {Sheikh-Jabbari}}]{Compare2016}%
  \BibitemOpen
  \bibfield  {author} {\bibinfo {author} {\bibfnamefont {G.}~\bibnamefont
  {Comp{\`e}re}}, \bibinfo {author} {\bibfnamefont {P.}~\bibnamefont {Mao}},
  \bibinfo {author} {\bibfnamefont {A.}~\bibnamefont {Seraj}}, \ and\ \bibinfo
  {author} {\bibfnamefont {M.~M.}\ \bibnamefont {Sheikh-Jabbari}},\ }\href
  {\doibase 10.1007/JHEP01(2016)080} {\bibfield  {journal} {\bibinfo  {journal}
  {Journal of High Energy Physics}\ }\textbf {\bibinfo {volume} {2016}},\
  \bibinfo {pages} {80} (\bibinfo {year} {2016})}\BibitemShut {NoStop}%
\bibitem [{\citenamefont {Wald}(2001)}]{Wald2001}%
  \BibitemOpen
  \bibfield  {author} {\bibinfo {author} {\bibfnamefont {R.~M.}\ \bibnamefont
  {Wald}},\ }\href {\doibase 10.12942/lrr-2001-6} {\bibfield  {journal}
  {\bibinfo  {journal} {Living Reviews in Relativity}\ }\textbf {\bibinfo
  {volume} {4}},\ \bibinfo {pages} {6} (\bibinfo {year} {2001})}\BibitemShut
  {NoStop}%
\bibitem [{\citenamefont {Gour}\ and\ \citenamefont
  {Mayo}(2001)}]{PhysRevD.63.064005}%
  \BibitemOpen
  \bibfield  {author} {\bibinfo {author} {\bibfnamefont {G.}~\bibnamefont
  {Gour}}\ and\ \bibinfo {author} {\bibfnamefont {A.~E.}\ \bibnamefont
  {Mayo}},\ }\href {\doibase 10.1103/PhysRevD.63.064005} {\bibfield  {journal}
  {\bibinfo  {journal} {Phys. Rev. D}\ }\textbf {\bibinfo {volume} {63}},\
  \bibinfo {pages} {064005} (\bibinfo {year} {2001})}\BibitemShut {NoStop}%
\bibitem [{\citenamefont {Fatibene}\ \emph {et~al.}(1999)\citenamefont
  {Fatibene}, \citenamefont {Ferraris}, \citenamefont {Francaviglia},\ and\
  \citenamefont {Raiteri}}]{Fatibene:1998rq}%
  \BibitemOpen
  \bibfield  {author} {\bibinfo {author} {\bibfnamefont {L.}~\bibnamefont
  {Fatibene}}, \bibinfo {author} {\bibfnamefont {M.}~\bibnamefont {Ferraris}},
  \bibinfo {author} {\bibfnamefont {M.}~\bibnamefont {Francaviglia}}, \ and\
  \bibinfo {author} {\bibfnamefont {M.}~\bibnamefont {Raiteri}},\ }\href
  {\doibase 10.1006/aphy.1999.5915} {\bibfield  {journal} {\bibinfo  {journal}
  {Annals Phys.}\ }\textbf {\bibinfo {volume} {275}},\ \bibinfo {pages} {27}
  (\bibinfo {year} {1999})},\ \Eprint {http://arxiv.org/abs/hep-th/9810039}
  {arXiv:hep-th/9810039 [hep-th]} \BibitemShut {NoStop}%
\bibitem [{\citenamefont {Carlip}(1999)}]{Carlip:1999cy}%
  \BibitemOpen
  \bibfield  {author} {\bibinfo {author} {\bibfnamefont {S.}~\bibnamefont
  {Carlip}},\ }\href {\doibase 10.1088/0264-9381/16/10/322} {\bibfield
  {journal} {\bibinfo  {journal} {Class. Quant. Grav.}\ }\textbf {\bibinfo
  {volume} {16}},\ \bibinfo {pages} {3327} (\bibinfo {year} {1999})},\ \Eprint
  {http://arxiv.org/abs/gr-qc/9906126} {arXiv:gr-qc/9906126 [gr-qc]}
  \BibitemShut {NoStop}%
\bibitem [{\citenamefont {Brustein}\ and\ \citenamefont
  {Gorbonos}(2009)}]{Brustein:2009wr}%
  \BibitemOpen
  \bibfield  {author} {\bibinfo {author} {\bibfnamefont {R.}~\bibnamefont
  {Brustein}}\ and\ \bibinfo {author} {\bibfnamefont {D.}~\bibnamefont
  {Gorbonos}},\ }\href {\doibase 10.1103/PhysRevD.79.126003} {\bibfield
  {journal} {\bibinfo  {journal} {Phys. Rev.}\ }\textbf {\bibinfo {volume}
  {D79}},\ \bibinfo {pages} {126003} (\bibinfo {year} {2009})},\ \Eprint
  {http://arxiv.org/abs/0902.1553} {arXiv:0902.1553 [hep-th]} \BibitemShut
  {NoStop}%
\bibitem [{\citenamefont {Aros}\ \emph {et~al.}(2010)\citenamefont {Aros},
  \citenamefont {Diaz},\ and\ \citenamefont {Montecinos}}]{Aros:2010jb}%
  \BibitemOpen
  \bibfield  {author} {\bibinfo {author} {\bibfnamefont {R.}~\bibnamefont
  {Aros}}, \bibinfo {author} {\bibfnamefont {D.~E.}\ \bibnamefont {Diaz}}, \
  and\ \bibinfo {author} {\bibfnamefont {A.}~\bibnamefont {Montecinos}},\
  }\href {\doibase 10.1007/JHEP07(2010)012} {\bibfield  {journal} {\bibinfo
  {journal} {JHEP}\ }\textbf {\bibinfo {volume} {07}},\ \bibinfo {pages} {012}
  (\bibinfo {year} {2010})},\ \Eprint {http://arxiv.org/abs/1003.1083}
  {arXiv:1003.1083 [hep-th]} \BibitemShut {NoStop}%
\bibitem [{\citenamefont {Majhi}\ and\ \citenamefont
  {Padmanabhan}(2012)}]{Majhi:2012tf}%
  \BibitemOpen
  \bibfield  {author} {\bibinfo {author} {\bibfnamefont {B.~R.}\ \bibnamefont
  {Majhi}}\ and\ \bibinfo {author} {\bibfnamefont {T.}~\bibnamefont
  {Padmanabhan}},\ }\href {\doibase 10.1103/PhysRevD.86.101501} {\bibfield
  {journal} {\bibinfo  {journal} {Phys. Rev.}\ }\textbf {\bibinfo {volume}
  {D86}},\ \bibinfo {pages} {101501} (\bibinfo {year} {2012})},\ \Eprint
  {http://arxiv.org/abs/1204.1422} {arXiv:1204.1422 [gr-qc]} \BibitemShut
  {NoStop}%
\bibitem [{\citenamefont {Chakraborty}\ and\ \citenamefont
  {Padmanabhan}(2015)}]{Chakraborty:2015hna}%
  \BibitemOpen
  \bibfield  {author} {\bibinfo {author} {\bibfnamefont {S.}~\bibnamefont
  {Chakraborty}}\ and\ \bibinfo {author} {\bibfnamefont {T.}~\bibnamefont
  {Padmanabhan}},\ }\href {\doibase 10.1103/PhysRevD.92.104011} {\bibfield
  {journal} {\bibinfo  {journal} {Phys. Rev.}\ }\textbf {\bibinfo {volume}
  {D92}},\ \bibinfo {pages} {104011} (\bibinfo {year} {2015})},\ \Eprint
  {http://arxiv.org/abs/1508.04060} {arXiv:1508.04060 [gr-qc]} \BibitemShut
  {NoStop}%
\bibitem [{\citenamefont {Setare}\ and\ \citenamefont
  {Adami}(2016)}]{Setare:2015nla}%
  \BibitemOpen
  \bibfield  {author} {\bibinfo {author} {\bibfnamefont {M.~R.}\ \bibnamefont
  {Setare}}\ and\ \bibinfo {author} {\bibfnamefont {H.}~\bibnamefont {Adami}},\
  }\href {\doibase 10.1016/j.nuclphysb.2015.11.018} {\bibfield  {journal}
  {\bibinfo  {journal} {Nucl. Phys.}\ }\textbf {\bibinfo {volume} {B902}},\
  \bibinfo {pages} {115} (\bibinfo {year} {2016})},\ \Eprint
  {http://arxiv.org/abs/1509.05972} {arXiv:1509.05972 [hep-th]} \BibitemShut
  {NoStop}%
\bibitem [{\citenamefont {Jacobson}\ and\ \citenamefont
  {Mohd}(2015)}]{Jacobson:2015uqa}%
  \BibitemOpen
  \bibfield  {author} {\bibinfo {author} {\bibfnamefont {T.}~\bibnamefont
  {Jacobson}}\ and\ \bibinfo {author} {\bibfnamefont {A.}~\bibnamefont
  {Mohd}},\ }\href {\doibase 10.1103/PhysRevD.92.124010} {\bibfield  {journal}
  {\bibinfo  {journal} {Phys. Rev.}\ }\textbf {\bibinfo {volume} {D92}},\
  \bibinfo {pages} {124010} (\bibinfo {year} {2015})},\ \Eprint
  {http://arxiv.org/abs/1507.01054} {arXiv:1507.01054 [gr-qc]} \BibitemShut
  {NoStop}%
\bibitem [{\citenamefont {York}(1972)}]{PhysRevLett.28.1082}%
  \BibitemOpen
  \bibfield  {author} {\bibinfo {author} {\bibfnamefont {J.~W.}\ \bibnamefont
  {York}},\ }\href {\doibase 10.1103/PhysRevLett.28.1082} {\bibfield  {journal}
  {\bibinfo  {journal} {Phys. Rev. Lett.}\ }\textbf {\bibinfo {volume} {28}},\
  \bibinfo {pages} {1082} (\bibinfo {year} {1972})}\BibitemShut {NoStop}%
\bibitem [{\citenamefont {Parattu}\ \emph
  {et~al.}(2016{\natexlab{a}})\citenamefont {Parattu}, \citenamefont
  {Chakraborty}, \citenamefont {Majhi},\ and\ \citenamefont
  {Padmanabhan}}]{Parattu:2015gga}%
  \BibitemOpen
  \bibfield  {author} {\bibinfo {author} {\bibfnamefont {K.}~\bibnamefont
  {Parattu}}, \bibinfo {author} {\bibfnamefont {S.}~\bibnamefont
  {Chakraborty}}, \bibinfo {author} {\bibfnamefont {B.~R.}\ \bibnamefont
  {Majhi}}, \ and\ \bibinfo {author} {\bibfnamefont {T.}~\bibnamefont
  {Padmanabhan}},\ }\href {\doibase 10.1007/s10714-016-2093-7} {\bibfield
  {journal} {\bibinfo  {journal} {Gen. Rel. Grav.}\ }\textbf {\bibinfo {volume}
  {48}},\ \bibinfo {pages} {94} (\bibinfo {year} {2016}{\natexlab{a}})},\
  \Eprint {http://arxiv.org/abs/1501.01053} {arXiv:1501.01053 [gr-qc]}
  \BibitemShut {NoStop}%
\bibitem [{\citenamefont {Parattu}\ \emph
  {et~al.}(2016{\natexlab{b}})\citenamefont {Parattu}, \citenamefont
  {Chakraborty},\ and\ \citenamefont {Padmanabhan}}]{Parattu:2016trq}%
  \BibitemOpen
  \bibfield  {author} {\bibinfo {author} {\bibfnamefont {K.}~\bibnamefont
  {Parattu}}, \bibinfo {author} {\bibfnamefont {S.}~\bibnamefont
  {Chakraborty}}, \ and\ \bibinfo {author} {\bibfnamefont {T.}~\bibnamefont
  {Padmanabhan}},\ }\href {\doibase 10.1140/epjc/s10052-016-3979-y} {\bibfield
  {journal} {\bibinfo  {journal} {Eur. Phys. J.}\ }\textbf {\bibinfo {volume}
  {C76}},\ \bibinfo {pages} {129} (\bibinfo {year} {2016}{\natexlab{b}})},\
  \Eprint {http://arxiv.org/abs/1602.07546} {arXiv:1602.07546 [gr-qc]}
  \BibitemShut {NoStop}%
\bibitem [{\citenamefont {Neiman}(2012)}]{Neiman:2012fx}%
  \BibitemOpen
  \bibfield  {author} {\bibinfo {author} {\bibfnamefont {Y.}~\bibnamefont
  {Neiman}},\ }\href@noop {} {\  (\bibinfo {year} {2012})},\ \Eprint
  {http://arxiv.org/abs/1212.2922} {arXiv:1212.2922 [hep-th]} \BibitemShut
  {NoStop}%
\bibitem [{\citenamefont {Neiman}(2013{\natexlab{a}})}]{Neiman:2013ap}%
  \BibitemOpen
  \bibfield  {author} {\bibinfo {author} {\bibfnamefont {Y.}~\bibnamefont
  {Neiman}},\ }\href {\doibase 10.1007/JHEP04(2013)071} {\bibfield  {journal}
  {\bibinfo  {journal} {JHEP}\ }\textbf {\bibinfo {volume} {04}},\ \bibinfo
  {pages} {071} (\bibinfo {year} {2013}{\natexlab{a}})},\ \Eprint
  {http://arxiv.org/abs/1301.7041} {arXiv:1301.7041 [gr-qc]} \BibitemShut
  {NoStop}%
\bibitem [{\citenamefont {Neiman}(2013{\natexlab{b}})}]{Neiman:2013lxa}%
  \BibitemOpen
  \bibfield  {author} {\bibinfo {author} {\bibfnamefont {Y.}~\bibnamefont
  {Neiman}},\ }\href {\doibase 10.1103/PhysRevD.88.024037} {\bibfield
  {journal} {\bibinfo  {journal} {Phys. Rev.}\ }\textbf {\bibinfo {volume}
  {D88}},\ \bibinfo {pages} {024037} (\bibinfo {year} {2013}{\natexlab{b}})},\
  \Eprint {http://arxiv.org/abs/1305.2207} {arXiv:1305.2207 [gr-qc]}
  \BibitemShut {NoStop}%
\bibitem [{\citenamefont {Jubb}\ \emph {et~al.}(2017)\citenamefont {Jubb},
  \citenamefont {Samuel}, \citenamefont {Sorkin},\ and\ \citenamefont
  {Surya}}]{Jubb:2016qzt}%
  \BibitemOpen
  \bibfield  {author} {\bibinfo {author} {\bibfnamefont {I.}~\bibnamefont
  {Jubb}}, \bibinfo {author} {\bibfnamefont {J.}~\bibnamefont {Samuel}},
  \bibinfo {author} {\bibfnamefont {R.}~\bibnamefont {Sorkin}}, \ and\ \bibinfo
  {author} {\bibfnamefont {S.}~\bibnamefont {Surya}},\ }\href {\doibase
  10.1088/1361-6382/aa6014} {\bibfield  {journal} {\bibinfo  {journal} {Class.
  Quant. Grav.}\ }\textbf {\bibinfo {volume} {34}},\ \bibinfo {pages} {065006}
  (\bibinfo {year} {2017})},\ \Eprint {http://arxiv.org/abs/1612.00149}
  {arXiv:1612.00149 [gr-qc]} \BibitemShut {NoStop}%
\bibitem [{\citenamefont {Lehner}\ \emph {et~al.}(2016)\citenamefont {Lehner},
  \citenamefont {Myers}, \citenamefont {Poisson},\ and\ \citenamefont
  {Sorkin}}]{Lehner:2016vdi}%
  \BibitemOpen
  \bibfield  {author} {\bibinfo {author} {\bibfnamefont {L.}~\bibnamefont
  {Lehner}}, \bibinfo {author} {\bibfnamefont {R.~C.}\ \bibnamefont {Myers}},
  \bibinfo {author} {\bibfnamefont {E.}~\bibnamefont {Poisson}}, \ and\
  \bibinfo {author} {\bibfnamefont {R.~D.}\ \bibnamefont {Sorkin}},\ }\href
  {\doibase 10.1103/PhysRevD.94.084046} {\bibfield  {journal} {\bibinfo
  {journal} {Phys. Rev.}\ }\textbf {\bibinfo {volume} {D94}},\ \bibinfo {pages}
  {084046} (\bibinfo {year} {2016})},\ \Eprint
  {http://arxiv.org/abs/1609.00207} {arXiv:1609.00207 [hep-th]} \BibitemShut
  {NoStop}%
\bibitem [{\citenamefont {Krishnan}\ and\ \citenamefont
  {Raju}(2017)}]{Krishnan:2016mcj}%
  \BibitemOpen
  \bibfield  {author} {\bibinfo {author} {\bibfnamefont {C.}~\bibnamefont
  {Krishnan}}\ and\ \bibinfo {author} {\bibfnamefont {A.}~\bibnamefont
  {Raju}},\ }\href {\doibase 10.1142/S0217732317500778} {\bibfield  {journal}
  {\bibinfo  {journal} {Mod. Phys. Lett.}\ }\textbf {\bibinfo {volume} {A32}},\
  \bibinfo {pages} {1750077} (\bibinfo {year} {2017})},\ \Eprint
  {http://arxiv.org/abs/1605.01603} {arXiv:1605.01603 [hep-th]} \BibitemShut
  {NoStop}%
\bibitem [{\citenamefont {Fr{\`e}}(2012)}]{fre2012gravity}%
  \BibitemOpen
  \bibfield  {author} {\bibinfo {author} {\bibfnamefont {P.}~\bibnamefont
  {Fr{\`e}}},\ }\href@noop {} {\emph {\bibinfo {title} {Gravity, a Geometrical
  Course: Volume 1: Development of the Theory and Basic Physical
  Applications}}},\ Gravity, a Geometrical Course\ (\bibinfo  {publisher}
  {Springer Netherlands},\ \bibinfo {year} {2012})\BibitemShut {NoStop}%
\bibitem [{\citenamefont {Ashtekar}(1988)}]{ashtekar1988new}%
  \BibitemOpen
  \bibfield  {author} {\bibinfo {author} {\bibfnamefont {A.}~\bibnamefont
  {Ashtekar}},\ }\href {https://books.google.co.jp/books?id=uYwzAQAAIAAJ}
  {\emph {\bibinfo {title} {New perspectives in canonical gravity}}},\
  Monographs and textbooks in physical science\ (\bibinfo  {publisher}
  {Bibliopolis},\ \bibinfo {year} {1988})\BibitemShut {NoStop}%
\bibitem [{\citenamefont {Carroll}(2004)}]{carroll2004spacetime}%
  \BibitemOpen
  \bibfield  {author} {\bibinfo {author} {\bibfnamefont {S.}~\bibnamefont
  {Carroll}},\ }\href {https://books.google.co.jp/books?id=1SKFQgAACAAJ} {\emph
  {\bibinfo {title} {Spacetime and Geometry: An Introduction to General
  Relativity}}}\ (\bibinfo  {publisher} {Addison Wesley},\ \bibinfo {year}
  {2004})\BibitemShut {NoStop}%
\bibitem [{\citenamefont {Padmanabhan}(2014)}]{Padmanabhan:2010zzb}%
  \BibitemOpen
  \bibfield  {author} {\bibinfo {author} {\bibfnamefont {T.}~\bibnamefont
  {Padmanabhan}},\ }\href {http://www.cambridge.org/9780521882231} {\emph
  {\bibinfo {title} {{Gravitation: Foundations and frontiers}}}}\ (\bibinfo
  {publisher} {Cambridge University Press},\ \bibinfo {year}
  {2014})\BibitemShut {NoStop}%
\bibitem [{\citenamefont {Bojowald}(2011)}]{bojowald2011quantum}%
  \BibitemOpen
  \bibfield  {author} {\bibinfo {author} {\bibfnamefont {M.}~\bibnamefont
  {Bojowald}},\ }\href {https://books.google.co.jp/books?id=CpYVdWkL\_ZYC}
  {\emph {\bibinfo {title} {Quantum Cosmology: A Fundamental Description of the
  Universe}}},\ Lecture Notes in Physics\ (\bibinfo  {publisher} {Springer New
  York},\ \bibinfo {year} {2011})\BibitemShut {NoStop}%
\bibitem [{\citenamefont {Rovelli}\ and\ \citenamefont
  {Vidotto}(2014)}]{rovelli2014covariant}%
  \BibitemOpen
  \bibfield  {author} {\bibinfo {author} {\bibfnamefont {C.}~\bibnamefont
  {Rovelli}}\ and\ \bibinfo {author} {\bibfnamefont {F.}~\bibnamefont
  {Vidotto}},\ }\href {https://books.google.co.jp/books?id=4VjeBAAAQBAJ} {\emph
  {\bibinfo {title} {Covariant Loop Quantum Gravity: An Elementary Introduction
  to Quantum Gravity and Spinfoam Theory}}},\ Cambridge Monographs on
  Mathematical Physics\ (\bibinfo  {publisher} {Cambridge University Press},\
  \bibinfo {year} {2014})\BibitemShut {NoStop}%
\bibitem [{\citenamefont {Poisson}\ and\ \citenamefont
  {Will}(2014)}]{poisson2014gravity}%
  \BibitemOpen
  \bibfield  {author} {\bibinfo {author} {\bibfnamefont {E.}~\bibnamefont
  {Poisson}}\ and\ \bibinfo {author} {\bibfnamefont {C.}~\bibnamefont {Will}},\
  }\href {https://books.google.co.jp/books?id=PZ5cAwAAQBAJ} {\emph {\bibinfo
  {title} {Gravity: Newtonian, Post-Newtonian, Relativistic}}}\ (\bibinfo
  {publisher} {Cambridge University Press},\ \bibinfo {year}
  {2014})\BibitemShut {NoStop}%
\bibitem [{\citenamefont {{Kurihara}}(2017)}]{2017arXiv170305574K}%
  \BibitemOpen
  \bibfield  {author} {\bibinfo {author} {\bibfnamefont {Y.}~\bibnamefont
  {{Kurihara}}},\ }\href@noop {} {\bibfield  {journal} {\bibinfo  {journal}
  {ArXiv e-prints}\ } (\bibinfo {year} {2017})},\ \Eprint
  {http://arxiv.org/abs/1703.05574} {arXiv:1703.05574 [physics.gen-ph]}
  \BibitemShut {NoStop}%
\bibitem [{\citenamefont {Kurihara}\ \emph {et~al.}(2014)\citenamefont
  {Kurihara}, \citenamefont {Phan},\ and\ \citenamefont
  {Quach}}]{kurihara2014}%
  \BibitemOpen
  \bibfield  {author} {\bibinfo {author} {\bibfnamefont {Y.}~\bibnamefont
  {Kurihara}}, \bibinfo {author} {\bibfnamefont {K.~H.}\ \bibnamefont {Phan}},
  \ and\ \bibinfo {author} {\bibfnamefont {N.~M.~U.}\ \bibnamefont {Quach}},\
  }\href@noop {} {\bibfield  {journal} {\bibinfo  {journal} {J. Theor. Appl.
  Phys.}\ }\textbf {\bibinfo {volume} {8}},\ \bibinfo {pages} {143} (\bibinfo
  {year} {2014})}\BibitemShut {NoStop}%
\bibitem [{Note1()}]{Note1}%
  \BibitemOpen
  \bibinfo {note} {The factor $c/\hbar $ was restored by the author to make the
  entropy dimensionless. This factor was not explicitly written (set to unity)
  in Eq. (26) in the original paper\cite {PhysRevD.48.R3427}.}\BibitemShut
  {Stop}%
\bibitem [{\citenamefont {Boyer}\ and\ \citenamefont
  {Lindquist}(1967)}]{BoyerLindquist}%
  \BibitemOpen
  \bibfield  {author} {\bibinfo {author} {\bibfnamefont {R.~D.}\ \bibnamefont
  {Boyer}}\ and\ \bibinfo {author} {\bibfnamefont {R.~W.}\ \bibnamefont
  {Lindquist}},\ }\href {\doibase 10.1063/1.1705193} {\bibfield  {journal}
  {\bibinfo  {journal} {J. Math. Phys}\ }\textbf {\bibinfo {volume} {8}},\
  \bibinfo {pages} {265} (\bibinfo {year} {1967})}\BibitemShut {NoStop}%
\bibitem [{\citenamefont {Kerr}(1963)}]{PhysRevLett.11.237}%
  \BibitemOpen
  \bibfield  {author} {\bibinfo {author} {\bibfnamefont {R.~P.}\ \bibnamefont
  {Kerr}},\ }\href {\doibase 10.1103/PhysRevLett.11.237} {\bibfield  {journal}
  {\bibinfo  {journal} {Phys. Rev. Lett.}\ }\textbf {\bibinfo {volume} {11}},\
  \bibinfo {pages} {237} (\bibinfo {year} {1963})}\BibitemShut {NoStop}%
\bibitem [{\citenamefont {Newman}\ \emph {et~al.}(1965)\citenamefont {Newman},
  \citenamefont {Couch}, \citenamefont {Chinnapared}, \citenamefont {Exton},
  \citenamefont {Prakash},\ and\ \citenamefont {Torrence}}]{Newman}%
  \BibitemOpen
  \bibfield  {author} {\bibinfo {author} {\bibfnamefont {E.~T.}\ \bibnamefont
  {Newman}}, \bibinfo {author} {\bibfnamefont {E.}~\bibnamefont {Couch}},
  \bibinfo {author} {\bibfnamefont {K.}~\bibnamefont {Chinnapared}}, \bibinfo
  {author} {\bibfnamefont {A.}~\bibnamefont {Exton}}, \bibinfo {author}
  {\bibfnamefont {A.}~\bibnamefont {Prakash}}, \ and\ \bibinfo {author}
  {\bibfnamefont {R.}~\bibnamefont {Torrence}},\ }\href {\doibase
  http://dx.doi.org/10.1063/1.1704351} {\bibfield  {journal} {\bibinfo
  {journal} {Journal of Mathematical Physics}\ }\textbf {\bibinfo {volume}
  {6}},\ \bibinfo {pages} {918} (\bibinfo {year} {1965})}\BibitemShut {NoStop}%
\bibitem [{\citenamefont {Gibbons}\ and\ \citenamefont
  {Hawking}(1977)}]{PhysRevD.15.2738}%
  \BibitemOpen
  \bibfield  {author} {\bibinfo {author} {\bibfnamefont {G.~W.}\ \bibnamefont
  {Gibbons}}\ and\ \bibinfo {author} {\bibfnamefont {S.~W.}\ \bibnamefont
  {Hawking}},\ }\href {\doibase 10.1103/PhysRevD.15.2738} {\bibfield  {journal}
  {\bibinfo  {journal} {Phys. Rev. D}\ }\textbf {\bibinfo {volume} {15}},\
  \bibinfo {pages} {2738} (\bibinfo {year} {1977})}\BibitemShut {NoStop}%
\bibitem [{\citenamefont {Caldarelli}\ \emph {et~al.}(2000)\citenamefont
  {Caldarelli}, \citenamefont {Cognola},\ and\ \citenamefont
  {Klemm}}]{0264-9381-17-2-310}%
  \BibitemOpen
  \bibfield  {author} {\bibinfo {author} {\bibfnamefont {M.~M.}\ \bibnamefont
  {Caldarelli}}, \bibinfo {author} {\bibfnamefont {G.}~\bibnamefont {Cognola}},
  \ and\ \bibinfo {author} {\bibfnamefont {D.}~\bibnamefont {Klemm}},\ }\href
  {http://stacks.iop.org/0264-9381/17/i=2/a=310} {\bibfield  {journal}
  {\bibinfo  {journal} {Classical and Quantum Gravity}\ }\textbf {\bibinfo
  {volume} {17}},\ \bibinfo {pages} {399} (\bibinfo {year} {2000})}\BibitemShut
  {NoStop}%
\bibitem [{Note2()}]{Note2}%
  \BibitemOpen
  \bibinfo {note} {See, for example, section $2.2$ in ref.\cite
  {fre2012gravity2}.}\BibitemShut {Stop}%
\bibitem [{\citenamefont {Einstein}(1922)}]{Einstein1922}%
  \BibitemOpen
  \bibfield  {author} {\bibinfo {author} {\bibfnamefont {A.}~\bibnamefont
  {Einstein}},\ }\href {\doibase 10.1007/BF01328424} {\bibfield  {journal}
  {\bibinfo  {journal} {Zeitschrift f\"ur Physik}\ }\textbf {\bibinfo {volume}
  {11}},\ \bibinfo {pages} {326} (\bibinfo {year} {1922})}\BibitemShut
  {NoStop}%
\bibitem [{\citenamefont {Lmea\^itre}(1927)}]{Lmeitre}%
  \BibitemOpen
  \bibfield  {author} {\bibinfo {author} {\bibfnamefont {G.}~\bibnamefont
  {Lmea\^itre}},\ }\href@noop {} {\bibfield  {journal} {\bibinfo  {journal}
  {Annals of the Scientific Society of Brussels}\ }\textbf {\bibinfo {volume}
  {A47}},\ \bibinfo {pages} {49} (\bibinfo {year} {1927})}\BibitemShut
  {NoStop}%
\bibitem [{\citenamefont {{Robertson}}(1935)}]{1935ApJ82284R}%
  \BibitemOpen
  \bibfield  {author} {\bibinfo {author} {\bibfnamefont {H.~P.}\ \bibnamefont
  {{Robertson}}},\ }\href {\doibase 10.1086/143681} {\bibfield  {journal}
  {\bibinfo  {journal} {The Astrophysical Journal}\ }\textbf {\bibinfo {volume}
  {82}},\ \bibinfo {pages} {284} (\bibinfo {year} {1935})}\BibitemShut
  {NoStop}%
\bibitem [{\citenamefont {Walker}(1937)}]{Walker01011937}%
  \BibitemOpen
  \bibfield  {author} {\bibinfo {author} {\bibfnamefont {A.~G.}\ \bibnamefont
  {Walker}},\ }\href {\doibase 10.1112/plms/s2-42.1.90} {\bibfield  {journal}
  {\bibinfo  {journal} {Proceedings of the London Mathematical Society}\
  }\textbf {\bibinfo {volume} {s2-42}},\ \bibinfo {pages} {90} (\bibinfo {year}
  {1937})},\ \Eprint
  {http://arxiv.org/abs/http://plms.oxfordjournals.org/content/s2-42/1/90.full.pdf+html}
  {http://plms.oxfordjournals.org/content/s2-42/1/90.full.pdf+html}
  \BibitemShut {NoStop}%
\end{thebibliography}%
\end{document}